# Image Acquisition System Using On Sensor Compressed Sampling Technique

**Pravir Singh Gupta**[a], **Gwan Seong Choi**[a]
[a]Texas A&M University , Department of Electrical and Computer Engineering, College Station, Texas, 77843

**Abstract.** Advances in CMOS technology have made high resolution image sensors possible. These image sensor pose significant challenges in terms of the amount of raw data generated, energy efficiency and frame rate. This paper presents a new design methodology for an imaging system and a simplified novel image sensor pixel design to be used in such system so that Compressed Sensing (CS) technique can be implemented easily at the sensor level. This results in significant energy savings as it not only cuts the raw data rate but also reduces transistor count per pixel, decreases pixel size, increases fill factor, simplifies ADC, JPEG encoder and JPEG decoder design and decreases wiring as well as address decoder size by half. Thus CS has the potential to increase the resolution of image sensors for a given technology and die size while significantly decreasing the power consumption and design complexity. We show that it has potential to reduce power consumption by about 23%-65%.

## 1 Introduction

In recent years the resolution of image sensors have increased at an amazing rate. Smartphones with 41 Mega-pixel cameras are available in the market. It is increasingly becoming difficult to handle the amount of data generated by such sensors in portable devices such as smartphones and cameras in terms of power requirements. If we use a byte of data (which is modest) to store the color of a pixel in RGB format we have 3 MB raw data per image for a 1 Mega-pixel camera. For a 41 Megapixel camera we have massive 123 MB raw data to process in hundreds of milliseconds. This poses a huge challenge given the power constraints of mobile devices and numerous snapshots and amount of data users are generating today in the multimedia-centric world. While we have huge secondary storage these days e.g. 128 GB SD/Micro-SD cards, the challenge is to handle the raw data generated at the sensor. Certainly, some sort of energy efficient modification has to be done in the traditional image acquisition system to handle the amount of data. If the compression is done at the sensor itself, we can avoid the huge bus wires, decrease clock rate and reduce the register widths. This will result in significant power savings as I/O read out will be reduced proportionately.

In the recent years, a lot of research has been conducted for compressively sampling of natural images. According to CS theory, if a signal is sparse in some domain, it can be recovered faithfully from a small number of linear combinations of the signal values provided that the matrix representing the linear combinations is incoherent with sparse domain basis vectors. But the traditional methods of CS makes matter worse when comes to acquisition effort per bit and storage effort per bit. Oike et al. (Ref. [1]) applied CS at Analog-to-Digital conversion level. The biggest issue with that approach is the sampled image looses image-like properties and hence image compression techniques like JPEG do not work well resulting in increase of storage effort per bit. Also, each pixel is read out multiple times which results in some wastage of energy and acquisition time. It also uses a pseudo-random generator which consumes additional energy. The design presented by Dadkhah et al. (Ref. [2]) does CS at the sensor level. But it wires the output of pseudo-random generator to each block. In addition to the problems associated with design presented by Oike et al. (Ref. [1]), it also consumes significant wiring area in the pixel and decreases the active area in the pixel. This will result in poor Peak-Signal-to-Noise-Ratio (PSNR) performance of pixel. Katic et al. (Ref. [3]) also present design on similar lines. Their design also contains random number generator which needs to be routed to pixels consuming wiring area and power. The goal of this paper is a very simplified implementation of CS such that it results in power savings, reduction in raw data rate, application of standard image compression techniques like JPEG post CS and simplification of hardware design while achieving optimal performance. To achieve this, the paper presents a new system design methodology for an imaging system and a simplified novel pixel design to be used in such system so that CS technique can be implemented easily at the sensor level. We show that pixels in our design flow are even simpler than the normal ones. In our paper, we have circumvented the need for pseudo-random generator by employing CS Super-Resolution technique presented by Sen et al. (Ref. [4]) with modifications. We present results from both binary permuted block diagonal sampling matrix as mentioned in Ref. [5] and Ref. [6] as well as our novel non-binary block diagonal sampling matrix. These are easy to implement in hardware and help us to perform on-sensor image compression. These matrix preserve image like properties so JPEG can be applied to compressively sampled image unlike traditional designs. We show that our design methodology has the potential to achieve 23%-65% power savings.

## 2 Background and Motivation

This section introduces the background concepts and motivation behind our system design methodology as well as pixel design.

### 2.1 *Compressed Sensing (CS) Theory*

Suppose we have signal $X$, having $N$ samples such that, $X \in R^{N\times 1}$. And we want to recover $X$ from $Y$, where

$$Y = \Phi X, \tag{1}$$

such that $\Phi$ is a $M \times N$ matrix and $M << N$. Because number of unknowns is significantly larger than observations, it is difficult to recover $X$ from $Y$ because Eq. (1) has infinitely many possible solutions. But if $X$ is sufficiently sparse, exact recovery is possible. This is compressed sensing (Ref. [7]). A popular choice for $\Phi$ i.e. measurement basis, is randomly generated matrix. In this work we also assume $\Phi$ is orthonormal i.e.

$$\Phi\Phi^T = I. \tag{2}$$

Lets say that the signal $X$ is sparse in some domain $\Psi$. Then the signal can be represented in sparse domain as follows -

$$X = \Psi T, \tag{3}$$

where $T$ represents the signal $X$ in the transform domain $\Psi^{-1}$. Using Eq. (3) and Eq. (1) we get,

$$Y = \Phi\Psi T. \tag{4}$$

Lets assume following,

$$A = \Phi\Psi. \tag{5}$$

Then using Eq. (4) and Eq. (5) we get,

$$Y = AT. \tag{6}$$

Since $A$ is $M \times N$ and $M << N$ recovery of the original signal is difficult because the system of equation represented by Eq. (6) has infinitely many solution. This is where CS comes to rescue. If the sensing matrix $A$ satisfies the Restricted Isometric Property stated (RIP) (Ref. [8]) below

$$1 - \epsilon \leq \frac{||AT||_2}{||T||_2} \leq 1 + \epsilon, \tag{7}$$

for some $\epsilon > 0$ then perfect reconstruction is guaranteed with very high probability. To reconstruct signal we can solve the following equation using linear programming techniques.

$$\min_{T} ||T||_{l1} \; such \; that \; Y = AT. \tag{8}$$

Another condition related to RIP is that sparsity basis should be incoherent with the sampling basis (Ref. [9]). The coherence between the two can be calculated as follows -

$$\mu(\Phi, \Psi) = \sqrt{N} \times max_{1\leq k,j\leq N}|\phi_k, \psi_j|, \tag{9}$$

where $\phi$ and $\psi$ are the basis vectors in sampling basis $\Phi$ and sparsity basis $\Psi$ respectively. The coherence ranges from $1$ to $\sqrt{N}$. If $\mu$ is close to $1$ then matrices are incoherent and vice versa. While the requirement of incoherence is implicit in Eq. (7), it is explicit in another sufficient condition for recovery of compressively sampled signals. Select $M$ measurements uniformly at random in $\Phi$ domain. Then if,

$$M > C\mu^2(\Phi, \Psi)S \log N, \tag{10}$$

for some positive constant $C$ and $S$-sparse signal (i.e. only S coefficients of signal are non-zero), solution to Eq. (8) is guaranteed with very high probability (Ref. [10]). Eq. (10) also indicates that if incoherence is less we need more samples to reconstruct the original signal with high probability (Ref. [9]).

The above discussion was applicable to strictly sparse signal which means the signal has a lot of perfect zero values when represented in sparse domain. But such signals are rarely found in nature. Images represented in the matrix form are no exception. Many natural signals are only approximately sparse, which means most of the coefficients are very small in magnitude. In such case, small coefficients can be discarded without much loss of perceptual quality. Lets say the signal $X$ is approximately sparse. Lets set all but $S$ largest elements of our approximately sparse signal $X$ as zero and denote the resulting signal by $X_S$. Lets denote the corresponding transform by $T_S$. Because $\Psi$ is orthonormal basis,

$$||T - T_S||_2 = ||X - X_S||_2. \tag{11}$$

So if $T$ can be classified as sparse or compressible, meaning sorted magnitudes of the $T$ decay quickly, then $X$ can be approximated by $X_S$ and, therefore, the error $||X - X_S||_2$ is small (Ref. [10]). This means we can discard a significant fraction of coefficients without much loss of quality. This is why CS works well with natural images.

For images popular sparsity basis are Wavelet, Fourier or Gradient. The measurement matrices which satisfy incoherence requirements broadly fall in 4 categories - random or Gaussian random matrices (Ref. [11]), scrambled Fourier matrices (Ref. [12]), Partial Noiselets (Ref. [9]) and scrambled block Hadamard matrices (Ref. [5, 13]). Unfortunately, these matrices have very expensive and challenging hardware implementation. Any attempt to implement these matrices negates the advantage gained by CS in terms of sampling effort per bit. To make matter worse, storage of the sampled image becomes even more challenging.

For images, the sampling matrix can be quite huge i.e. of the order of 1 Million. Storing or generating a matrix of such size is not feasible in a camera or a portable device. To solve this problem Block-Based CS is used which is explained in next subsection.

### 2.2 Block-Based CS

In block-based CS sampling, the image is divided into $B \times B$ blocks. The sampling is done using $\frac{M}{N} \times B^2$ sampling matrix where compression ratio = $\frac{M}{N}$. Hence we need to store only $\frac{M}{N}B \times B$ numbers rather than the full ensemble which results in huge savings in circuitry and power (Ref. [14]).

$$\Phi = \begin{pmatrix} \phi_B & & & & & \\ & \phi_B & & & & \\ & & \phi_B & & & \\ & & & \cdot & & \\ & & & & \cdot & \\ & & & & & \phi_B \end{pmatrix} \tag{12}$$

where, off-diagonal elements are all zeros.

For block-based CS image has to be vectorized in one dimension either by using raster scan or by just reshaping the matrix. There is a trade-off involved between memory and reconstruction performance in the selection of block dimension. Small $B$ means less memory but poor reconstruction performance while large $B$ means more memory but superior reconstruction performance. Here we have used an even simplified version of block CS. We have not vectorized the image in one dimension. Instead, we keep the image as such and use $(M/N \times B) \times B$ sampling matrix. This leads to even more simplified implementation. For our case, the block size does not have any affect on reconstruction performance in our simulation. An explanation about this has been provided in Sec. 3. Hence we choose smallest possible block size (i.e. $2 \times 4$) for simplicity.

The next subsection introduces the transform domain in which natural images are sparse, a key

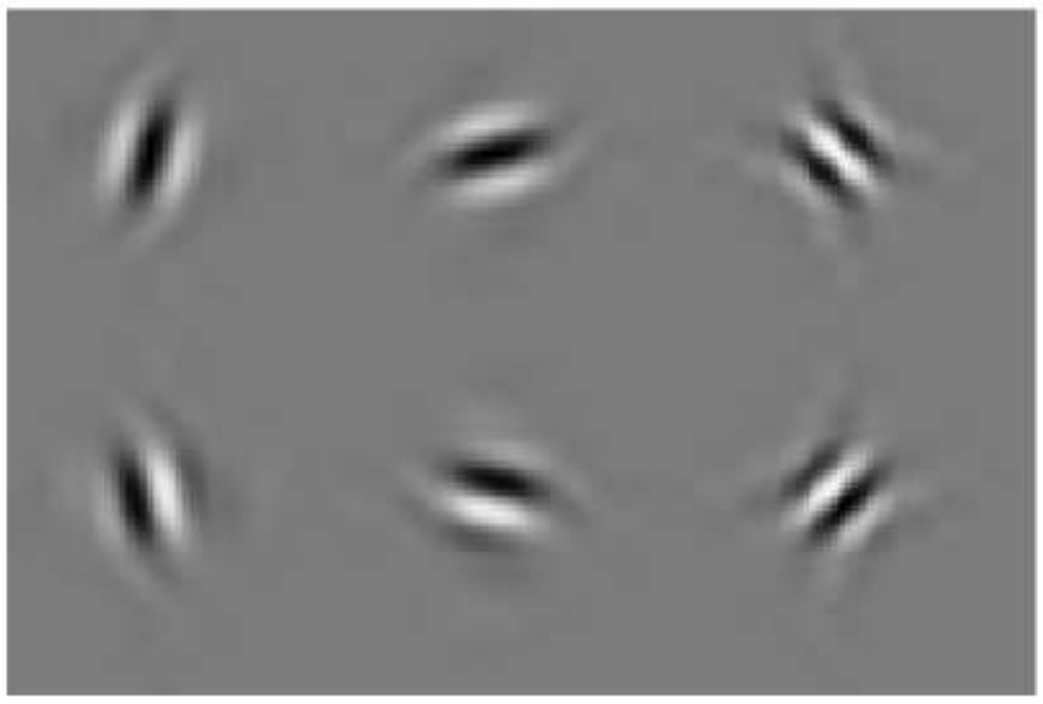

**Fig 1** The six wavelets.

requirement for compressed sensing/reconstruction.

### *2.3 Directional Transforms for Sparse Representation*

There are many transforms which can be used to represent an image as a sparse or approximately sparse signal. A popular one is Discrete Wavelet Transform (DWT). DWT lacks important properties such as shift invariance or directional selectivity. There are many modifications to DWT which have been extensively studied to preserve a much higher degree of directional representation than DWTs. One of them is DDWT (Dual Tree Discrete Wavelet Transform) (Ref. [15]). DDWT has an advantage over DWT as it provides efficient representation of directional features such as edges and contours. It has a redundancy of $2^m : 1$ for m-dimensional signals. Hence for 2-dimensional image, redundancy will be 4:1. It consists of both real and imaginary part but only real or imaginary part of DDWT guarantees perfect reconstruction and hence can be used as a standalone transform (Ref. [16]). While DWT is ambiguous in directionality property, mixing $+45$ and $-45$ together, DDWT has unique wavelet in each direction. They are oriented at $+/-75, +/-15, +/-45$. The wavelets are shown in Fig. 1.

The next subsection introduces the reconstruction algorithms for images sampled using CS technique.

*2.4 Reconstruction Algorithm*

A major problem associated with Block based CS is blocking-artifacts. A solution to this problem was presented by Gan et al. (Ref. [14]) by incorporating Weiner filtering into the basic PL (Projected Landweber) framework. This filtering helps to impose smoothness as well as sparsity inherent in PL algorithm. The algorithm (Ref. [17]) is given below

$function\ x^{i+1} = SPL(X^i, y, \phi_B, \Psi, \lambda)$

$\hat{X}^i = Wiener(X^i)$

$for\ each\ block\ j$

$\hat{\hat{X}}_j^i = \hat{X}_j^i + \phi_B^T(y - \phi_B \hat{X}_j^i)$

$\check{\check{T}}^i = \Psi^{-1}\hat{\hat{X}}$

$\check{T}^i = Threshold(\check{\check{T}}^i, \lambda)$

$\bar{X}^i = \Psi \check{T}^i$

$for\ each\ block\ j$

$X^{i+1} = \bar{X}_j^i + \phi_B^T(y - \phi_B \bar{X}_j^i).$

In the above algorithm Weiner() represents pixel-wise adaptive weiner filtering using a neighborhood of $3 \times 3$. The initial value is given below:

$$x^0 = \Phi^T y, \tag{13}$$

and the termination criteria is as follows -

$$|D^{(i+1)} - D^{(i)}| < 10^{-4}, \tag{14}$$

$$where,\ D^{(i)} = \frac{1}{\sqrt{N}}||x^i - \hat{\hat{x}}^{(i-1)}||_2. \quad (15)$$

The above sections were about compressed sensing and reconstruction. The next subsection introduces a popular image storage technique which is a key component in our system flow.

### *2.5 JPEG Theory*

JPEG stands for Joint Photographics Expert Group. It is a very widely used lossy image compression technique. It can perform both lossless and lossy compression, though lossy compression is very widely used mode of compression. Lossy compression relies on the fact that most of the image information is contained in very few coefficients in the Discrete Cosine Transform (DCT) domain. So a vast majority of insignificant coefficients can be discarded without much loss in perceptual quality resulting in large compression ratios.

JPEG first divides the image into $8 \times 8$ pixel blocks and then calculates DCT of each block. A quantizer rounds off the resulting DCT coefficients according to the quantization matrix, which controls the amount of compression one wants to do. This step represents ”lossy” part of JPEG but allows for large compression ratios. We can also control the amount of compression by appropriately setting the quantization matrix. After quantization, data is compressed further by the use of variable length encoding of these coefficients. While JPEG has been applied previously to CS sampled images (Ref. [18]) but compression performance has not been mentioned. Li et. al. (Ref. [18]) also use the Gaussian random matrix to compressively sample the image. When we sample an image with the Gaussian random matrix, the sampled image has Gaussian distribution and the image-like properties are lost. This results in a very poor JPEG compression performance which will significantly increase the effort/energy required to store the image.

### 2.6 *Deterministic CS and Super-Resolution (SR)*

Traditionally, the projection or sampling matrix $\Phi$ is chosen as Gaussian Random matrix as it possess good RIP and is highly incoherent with most sparsifying basis. However, hardware implementation of Gaussian random matrix is infeasible. A deterministic construction of sampling matrix can result in considerable simplification of hardware implementation. A method for deterministic construction of matrices were first introduced in detail in Ref. [19]. The author used finite fields to construct cyclic matrices which satisfy RIP. This is popularly known as deterministic CS. Other methods for deterministic construction have also been proposed such as one in Ref. [20] where authors used Euler Square based binary CS matrices which outperformed their Gaussian counterparts.

Super-resolution (SR) implies construction of high-resolution images from one or more low resolution images. Traditionally SR had been done using a set of low-resolution images. The idea is to enforce the constraint of sparsity in the transform domain such as wavelet to reconstruct the image. But using CS for SR means, that sampling matrix is no longer random but deterministic. The sampling or projection matrix for SR is guided by imaging model. SR sampling matrix $L$ can be viewed as product of two matrices as follows (see Ref. [21] ) -

$$L = R \times L_p, \tag{16}$$

where $R$ is decimation operator or downsampler and $L_p$ is low pass filter. Since there is a low pass filter involved in construction of $L$, it will have frequency discriminative nature. It will filter out high frequency components but preserve low frequency components. Where as, a Gaussian random matrix will preserve all frequencies. This means $L$ exhibits good RIP characteristics for a class of signals that contain low frequency information only, but Gaussian random matrix has good characteristics for any class of signals (see Ref. [21]). However, in case of natural images, most of the energy in concentrated in low frequency signals only. Hence if cutoff frequency for $L_p$ is appropriately set, the loss might not be too much resulting in reasonable reconstruction. Lossy image compression algorithm too weed out or reduce the high frequency component while the process of compression. Sen et al. performed SR CS reconstruction (Ref. [4]) using filtered and point down sampled image. In our work, we present a novel image sensor design (see Sec. 4) for filtering and downsampling the image in the CMOS image sensor itself without additional hardware and resulting in significant power savings. An advantage is that because we are using filtering and downsampling, we do not need randomization of sampling matrix. This also results in significant savings in terms of hardware and power consumption as there is no need of random generator and associated wiring.

Next subsections will introduce the hardware aspect of image sensors.

### *2.7 Photodetectors*

There are mainly 3 types of photo sensing elements - photogates, phototransistors and photodiodes. In this work, we have used photodiodes. There are different types of photodiodes too. We have used simple p-n junction, although we can use more sophisticated p-i-n junction to improve the efficiency of an image sensor. As the name implies, p-i-n junction consists of intrinsic region between p and n region. The p-i-n junction device reduces dark current and charge-transfer noise (Ref. [22]). Hence using p-n junction over p-i-n junction does not affect the demonstration of main functionality of our system design methodology.

There are various types of p-n junction photodiodes also. They are - n+/p-sub, n-well/p-sub,

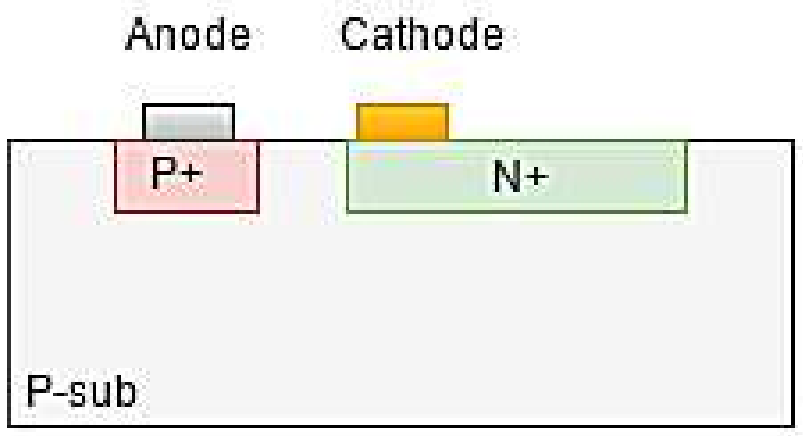

**Fig 2** n+/p-sub photodiode (Ref. [24]).

p+/n-well/p-sub. Murari el al. (Ref. [23]) list the parameters and advantages of various photodiodes. We are using n+/p-sub because of the large fill factor, low dark current per unit area values and ease of implementation to demonstrate our concept. Its schematic diagram is shown in Fig. 2.

### *2.8 Image Sensors*

In the past decade, extensive research has been done on CMOS sensors. An image pixel can be broadly divided into two parts, photo-detector element, and sensing circuit. Depending on sensing circuit there are two main families of image pixels, active pixel sensor, and passive pixel sensor. Passive pixel sensor carries out the charge of the photodetector and amplifies them later. Active pixel sensor has a photodetector and an active amplifier. Passive pixel sensors have mostly been implemented with Charge Coupled Device (CCD) technology while active pixel sensors are implemented using CMOS technology. Decreasing size and cost of CMOS elements has made CMOS image sensors viable and technology of choice (Ref. [25]). Ever decreasing size of transistors has made high-resolution image sensors possible. The most popular active pixel sensors design are 3T, 4T and CTIA (Capacitive Trans-Impedance Amplifier) pixels.

CTIA is mostly used in scientific applications while 3T and 4T are mostly used in commercial systems. We will not be discussing CTIA but the results presented can be applied in CTIA pixel as well. The schematic diagram for 3T and 4T pixel is shown in Fig. 3.

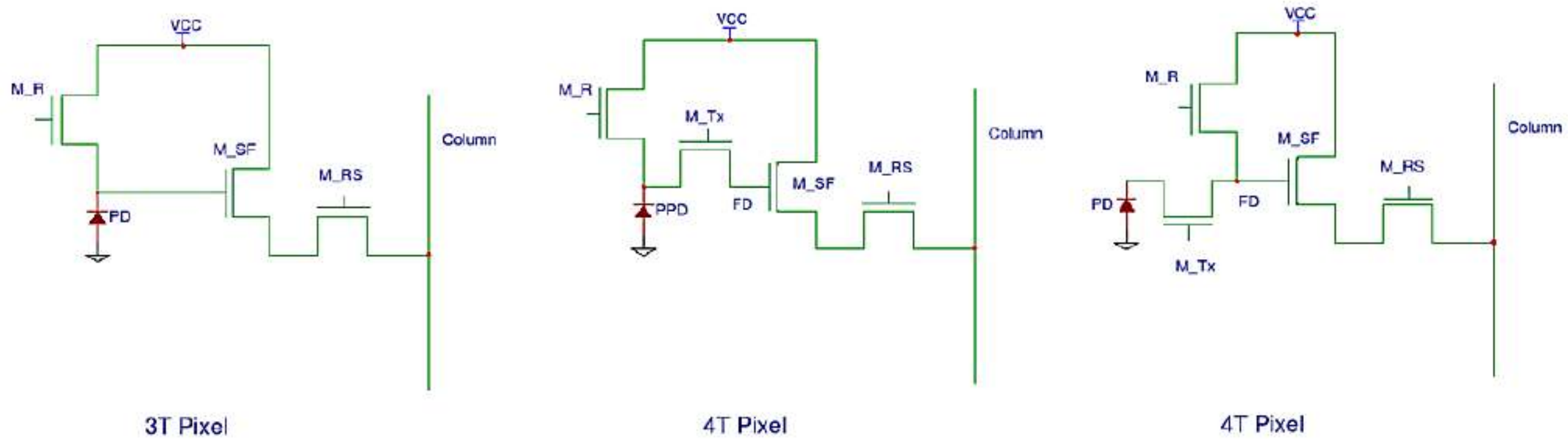

**Fig 3** 3T and 4T Pixel Schematic diagram. M_R stands for reset transistor, M_Tx stands for transmission gate, M_SF stands for source follower, M_RS stands for row select transistor, PD stands for photodiode, PPD stands for pinned photodiode and FD stands for floating diffusion node.

3T pixel is very compact but has less sensitivity and unstable bias voltage across photodiode. This pixel architecture consists of a photodiode and three transistors- Reset (M_R), Source Follower (M_SF) and a Row Select Transistor (M_RS). In 3T pixel operation, first the photodiode is reset using Reset transistor. Now, the charge gets collected on the photodiode proportional to light signal and exposure time. After a set integration time, the row select transistor is turned on to read out the signal using external readout circuitry.

The 4T (four transistor) pixel architecture is shown in Fig. 3 (Ref. [26]). Its architecture has two additional elements compared to the 3T architecture namely, the transfer gate (TX) and the floating diffusion node (FD). It uses either a Pinned Photo-Diode (PPD) or a normal Photo-Diode (PD) depending upon the design shown in Fig. 3. As long as TX is off, charge is accumulated in PPD or PD. When TX is on for set Integration time period, charge is transferred to the diffusion node. We have used 4T pixel design with PD as our choice for implementation as we did not have pinned photodiode (PPD) model to perform the simulation. It is expected that result will be similar with PPD as explained earlier subsection.

Because charge collection area and readout area are separated in the 4T pixel via M_Tx tran-sistor, it offers some key advantages. While the 3T design can only implement rolling shutter, the 4T design can implement both rolling as well as global shutter. Global shutter is very important for the high speed imaging application. The 4T pixel also allows low noise operation through the use of the Correlated Double Sampling (CDS) technique. The reset noise or kTC noise is the main source of noise resulting from the resetting operation of floating diffusion node through the resistive channel of the reset transistor. Thus, CDS technique can be employed to sample the floating diffusion node before and after M_Tx is turned on within a short time interval, thereby eliminating kTC noise. This operation is shown in Fig. 4.

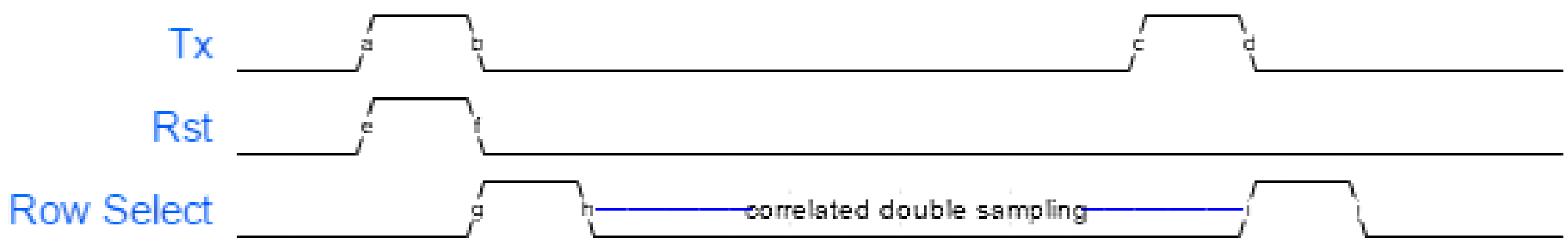

**Fig 4** Correlated Double Sampling (CDS) for a single image.

Transfer transistor or M_Tx makes the bias voltage across photodiode very stable. It also helps us to increase sensitivity because the integration capacitor can be kept small. CTIA has around 8 transistors but has highest sensitivity among all of them and stable photodiode voltage. Because of large pixel size, it is not much used in commercial systems. It is mostly used in scientific applications.

### *2.9 Nonidealities in Image Sensors*

Non-idealities can be broadly classified into two major groups - pixel level non-idealities and readout-level non-idealities (Ref. [3]). Both of them present challenges to the image sensor designers. Major pixel level non-idealities are - Dark Signal Non-Uniformity, Offset Fixed Pattern Noise, Photo-response non-uniformity, Pixel response non-linearity and pixel temporal noise. Ma-jor readout level non-linearities are - offset column fixed pattern noise, gain error column fixed pattern noise, readout non-linearity, readout temporal noise, readout output voltage range and quantization.

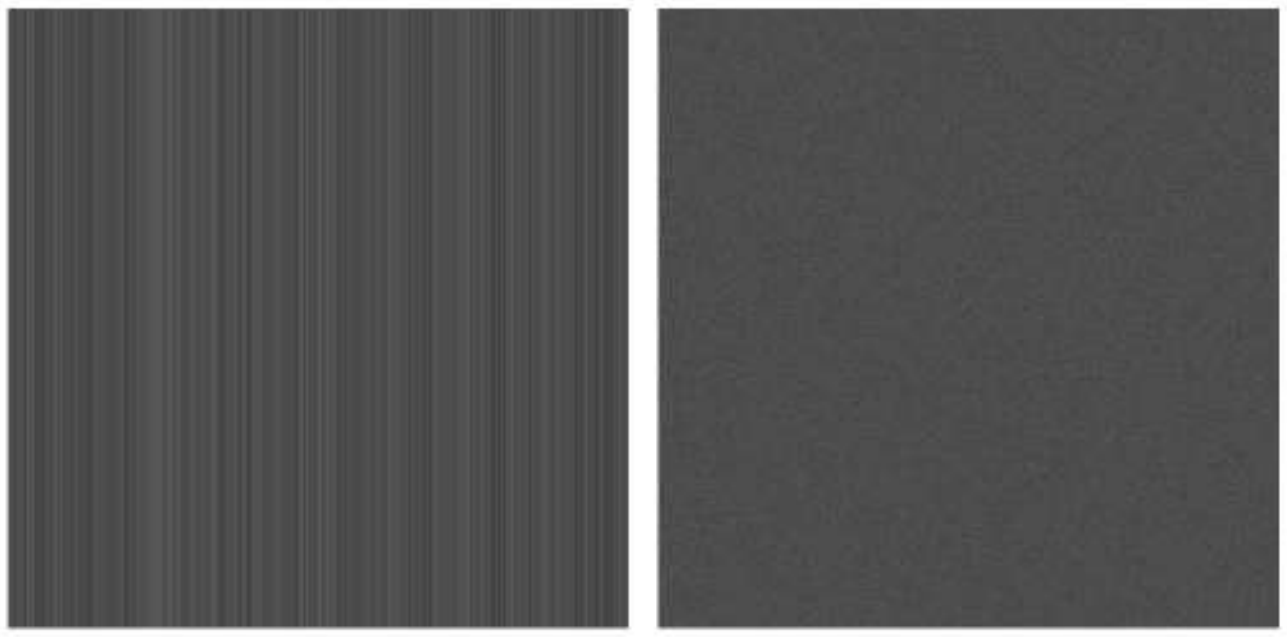

**Fig 5** Column and Pixel fixed pattern noise example.

In this paper, we will not deal with temporal noise but we will consider the fixed pattern noise and application of CS to overcome the challenges posed by fixed pattern noise. We are also not dealing with readout output voltage range and quantization noise as it is a research problem by itself and has been included in future course of our work. Offset fixed pattern noise can be easily dealt with by using correlated double sampling technique. But photoresponse non-uniformity, pixel response non-linearity and gain error fixed pattern noise require sophisticated circuitry to deal with. A simple way to deal with this problem is discussed in the Sec. 3 of the paper. An example image with column and pixel level fixed pattern noise is shown in Fig. 5.

## 3 Simulation

Our entire novel system design methodology can be described using the block diagram in Fig. 6. The input to the system is an image. The image gets sampled using compressed sampling technique using either of the sampling matrices. This sampling function is implemented in the image sensor itself. The design of such image sensor is discussed in Sec. 4 of this paper. Depending upon the quality desired, a desired number of bits are truncated while sampling. From here, the image goes to the image processor of the camera system. Here the image is compressed using JPEG technique as mentioned earlier. JPEG encoding can be done using ASIC chip or FPGA as well. We have used different levels of compression in JPEG to study the performance of our system which ranges from different levels of lossy to lossless compression. After compression, the image may get transmitted over a communication medium. The compressed image is then uncompressed. If the compression was lossy, there will be some loss of information. This uncompressed image is then reconstructed using SPL algorithm mentioned previously. The reconstruction performance is measured using PSNR metric.

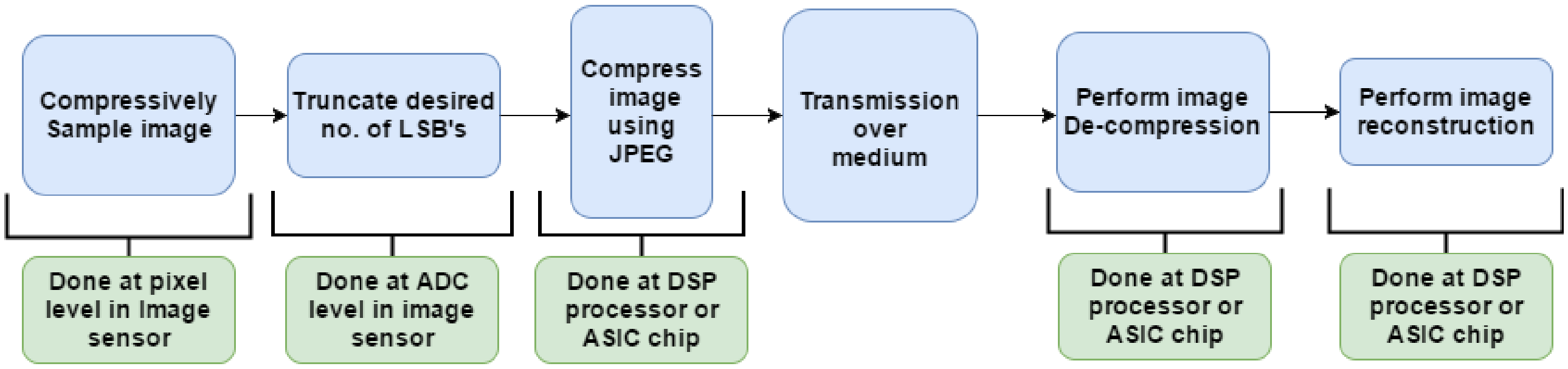

**Fig 6** Block diagram of system.

Now we demonstrate the reconstruction results of our proposed novel system flow. We use both binary and non-binary block diagonal matrix to compressively sample the image. The binary block diagonal ($\Phi_B$) and non-binary block diagonal ($\Phi_{NB}$) sampling matrix are mentioned below.

$$\Phi_B = \begin{pmatrix} 1 & 1 & 0 & 0 \\ 0 & 0 & 1 & 1 \end{pmatrix} \tag{17}$$

$$\Phi_{NB} = \begin{pmatrix} 9 & 7 & 0 & 0 \\ 0 & 0 & 9 & 7 \end{pmatrix} \tag{18}$$

Since matrix $\Phi_B$ adds two neighboring pixels, it does not significantly alter the statistical distribution of image and hence preserves image-like properties. But matrix $\Phi_{NB}$ performs weighted addition, so it does alter the distribution but still preserves some image-like properties. It does not alter the image as significantly as random Gaussian sampling matrix which makes the distribution of resulting sampled image as Gaussian. We arrived at $\Phi_{NB}$ empirically and it was found to be most optimal. One pixel is weighed approximately $1.3$ times relative to the other in $\Phi_{NB}$. One can try higher relative weights also but it will be difficult to implement in hardware due to large capacitor requirement (as per out design presented in next section). Since we have fixed bitwidth ADC's, we can only use integer weights to sample image otherwise we will loose the information contained in the decimal part. The image sampled using $\Phi_{NB}$ requires $12$ bits to store each pixel of resulting image. For $\Phi_B$ $9$ bits are required for the same.

Because of the way our sampling matrix is constructed, block size will not have any affect on reconstruction performance. We can see this from two matrices of different block sizes presented below.

$$\Phi_{B,2\times 4} = \begin{pmatrix} 1 & 1 & 0 & 0 \\ 0 & 0 & 1 & 1 \end{pmatrix} \tag{19}$$

$$\Phi_{B,4\times 8} = \begin{pmatrix} 1 & 1 & 0 & 0 & 0 & 0 & 0 & 0 \\ 0 & 0 & 1 & 1 & 0 & 0 & 0 & 0 \\ 0 & 0 & 0 & 0 & 1 & 1 & 0 & 0 \\ 0 & 0 & 0 & 0 & 0 & 0 & 1 & 1 \end{pmatrix} \tag{20}$$

We can see that both the above block matrix, when used as sampling matrix actually perform the same function of adding two rows. $4 \times 8$ matrix is actually two $2 \times 4$ matrix along the main diagonal of the sampling matrix presented in Eq. 12. Thus $4 \times 8$ can be expressed in terms of $2 \times 4$ matrix as follows -

$$\Phi_{B,4\times8} = \begin{pmatrix} \Phi_{B,2\times4} & \\ & \Phi_{B,2\times4} \end{pmatrix}. \tag{21}$$

Thus both of them lead to same sampling matrix presented in Eq. 12. Hence there will not be any affect in performance. The same reasoning applies to non-binary sampling matrix also. According to Eq. (16), Eq. (19) can be viewed as product of downsampler ($R$) and a circulant averaging filter ($L_p$). These matrices are as follows -

$$L_p = \begin{pmatrix} 1 & 1 & 0 & 0 \\ 0 & 1 & 1 & 0 \\ 0 & 0 & 1 & 1 \\ 1 & 0 & 0 & 1 \end{pmatrix} \tag{22}$$

$$R = \begin{pmatrix} 1 & 0 & 0 & 0 \\ 0 & 0 & 1 & 0 \end{pmatrix} \tag{23}$$

Similar lines of reasoning hold for Eq. (18).

Our sampling matrix in Eq. (17) and Eq. (18) is sparse leading to less incoherency with sparse transforms such as wavelet transform which is used as sparse basis (Ref. [5]). But it still works well for CS because according to Eq. (10) and Ref. [5] if incoherence is less, we need more samples to reconstruct signal with high probability. Since we are using $50\%$ compression in CS, this matrix works well as shown by our experimental results. Our sampling matrix in Eq. (17) and Eq. (18) do not satisfy Eq. (2) also. We call our sampling matrix in Eq. (17) and Eq. (18) as front-end sampling matrix. We have to perform a transformation on front-end sampling matrix so that it satisfies Eq. (2). The matrix resulting from transformation is known as back-end sampling matrix. We use front-end sampling matrix because it is very easy to implement on sensor level. The transformation from front-end to back-end is very simple. We have to just multiply the front-end sampling matrix by a normalization constant. The normalization constant is simply the square root of the sum of squares of all the elements in a row of the matrix.

$$\Phi_B = \frac{1}{\sqrt{N}} \times \Phi_F \tag{24}$$

where, $N$=$Sum\ of\ squares\ of\ row\ elements\ of\ matrix.$

The back end sampling matrix generated from Eq. (24) will satisfy Eq. (2) and this transformation can be implemented in the reconstruction algorithm itself. Multiplication of this transformation constant with the compressively sampled image (using front end sampling matrix) is equivalent to sampling the image using back-end sampling matrix which is what is desired. Thus we use back-end sampling matrix as the sampling matrix in the reconstruction algorithm. Using the transformation we calculated our back-end sampling matrix as follows -

$$\Phi_{B,backend} = \begin{pmatrix} 1/\sqrt{2} & 1/\sqrt{2} & 0 & 0 \\ 0 & 0 & 1/\sqrt{2} & 1/\sqrt{2} \end{pmatrix} \tag{25}$$

$$\Phi_{NB,backend} = \begin{pmatrix} .7894 & .6139 & 0 & 0 \\ 0 & 0 & .7894 & .6139 \end{pmatrix} \tag{26}$$

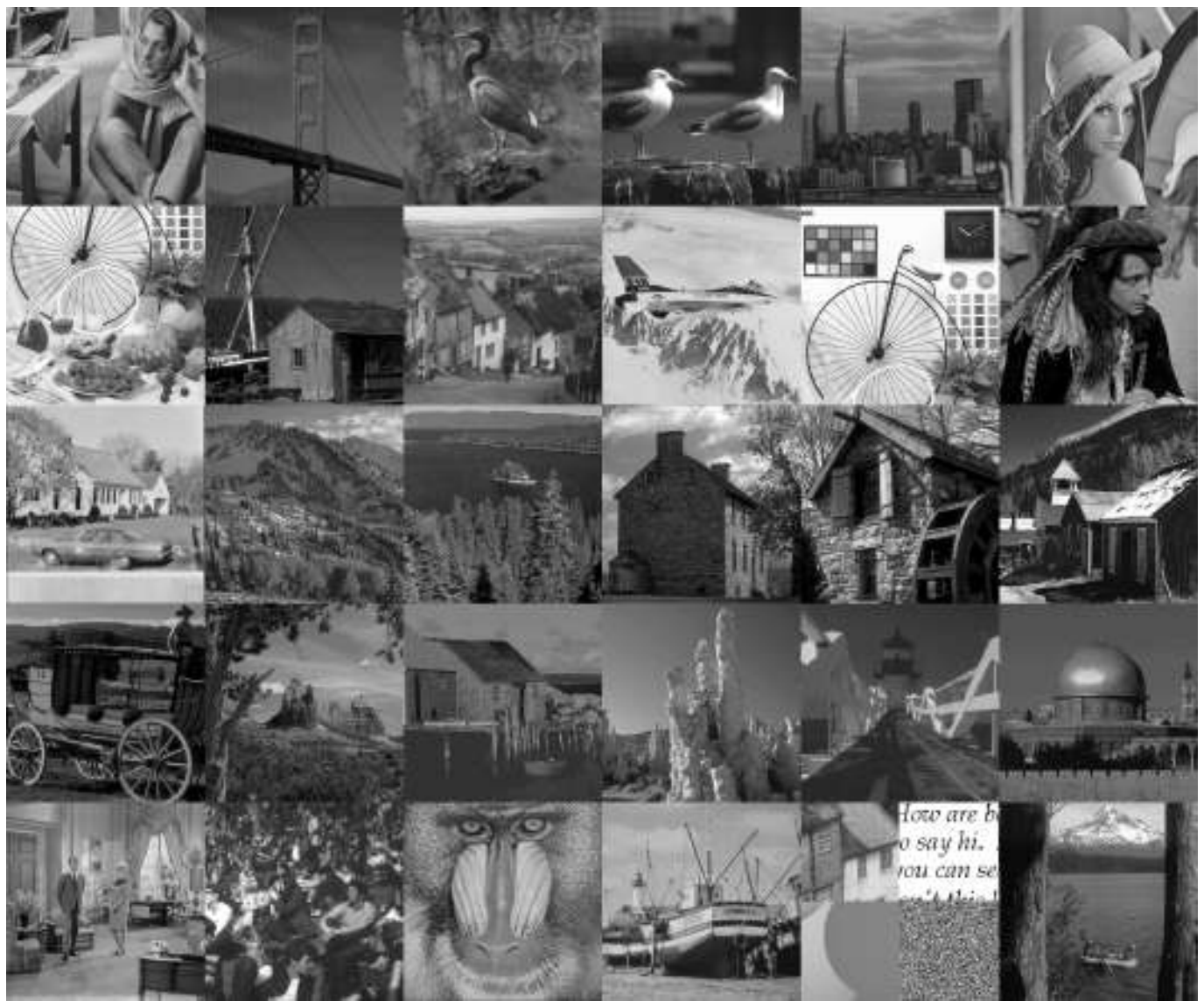

**Fig 7** The set of 30 images.

The level of lossy compression in JPEG is controlled using the $quality$ parameter of MATLAB function. We have measured the size of this compressed image. The baseline for image size is taken as the size of JPEG image ($quality = 75$, $bits = 8$), i.e. raw image stored as JPEG with $quality = 75$ and $bitdepth = 8$. The size of a compressively sampled image is reported as the relative percentage of this baseline. The baseline image for PSNR measurement is the raw image. The metrics for baseline is shown in Table 1.

We have used a set of $30$ images to perform the above simulation. These images are shown in Fig. 7. All the images are in Grayscale $512 \times 512$ format. In Fig. 8 we have shown how the average of the normalized size of the raw image stored in JPEG format scales with the Quality factor. Similarly in Fig. 9 we have shown how reconstruction performance of JPEG (measured as the average of PSNR of 30 images) varies with the Quality factor of JPEG.

The Table 2 lists the results for system depicted in Fig. 6 for binary sampling matrix with input parameters such as Quality factor of JPEG and Bitdepth of the image. The output values are

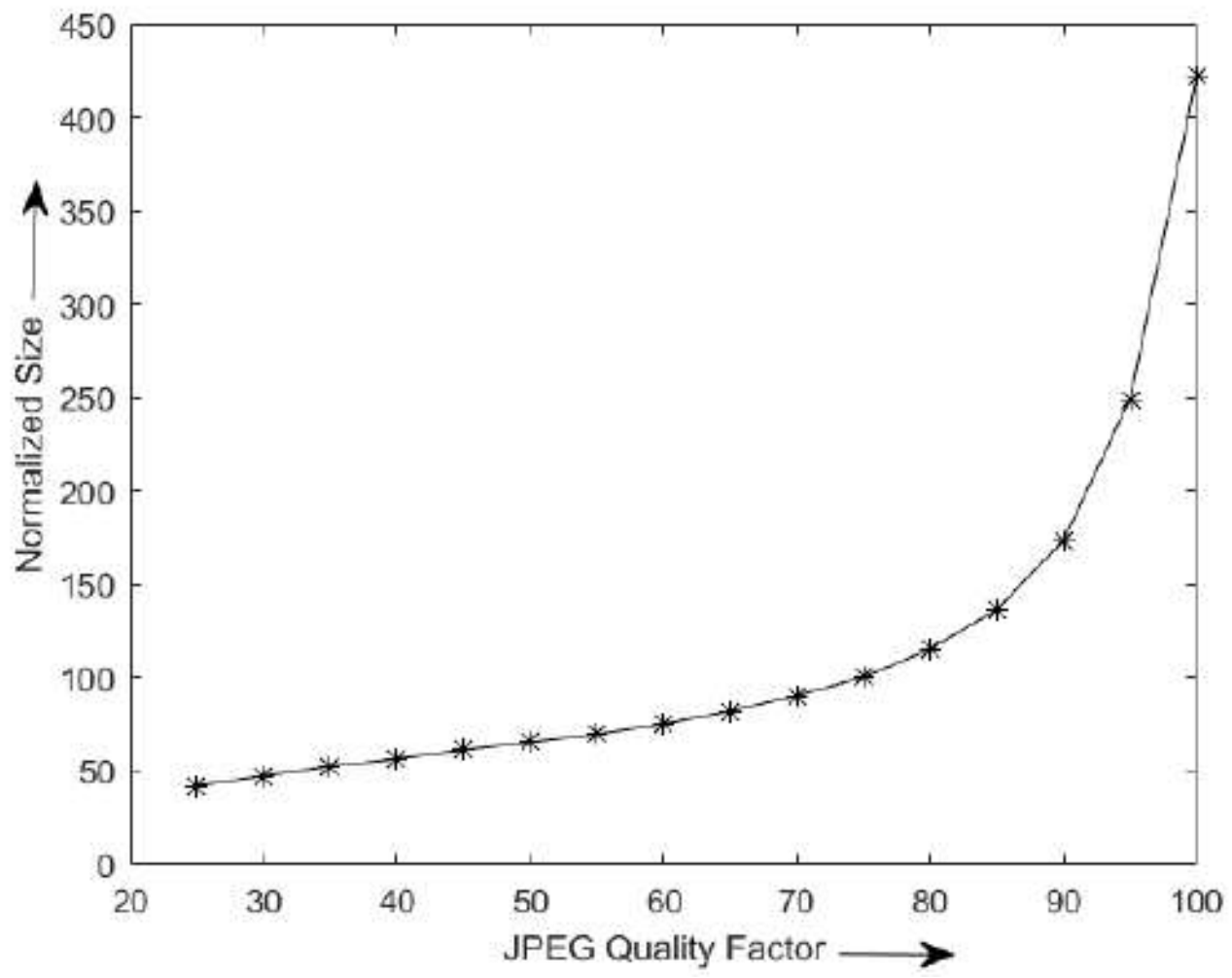

**Fig 8** Variation of Normalized Image Size with JPEG Quality factor.

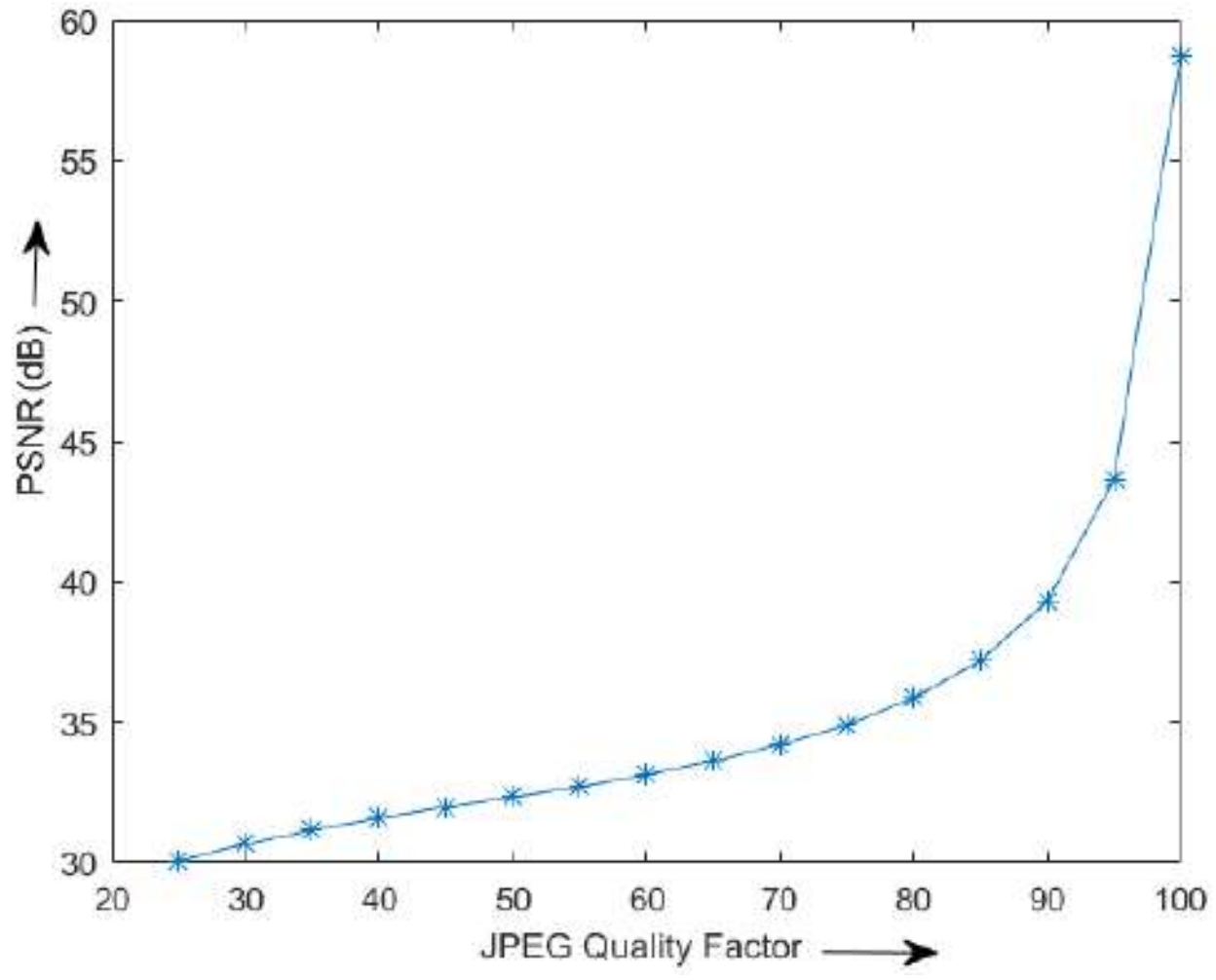

**Fig 9** Variation of PSNR with JPEG Quality factor.

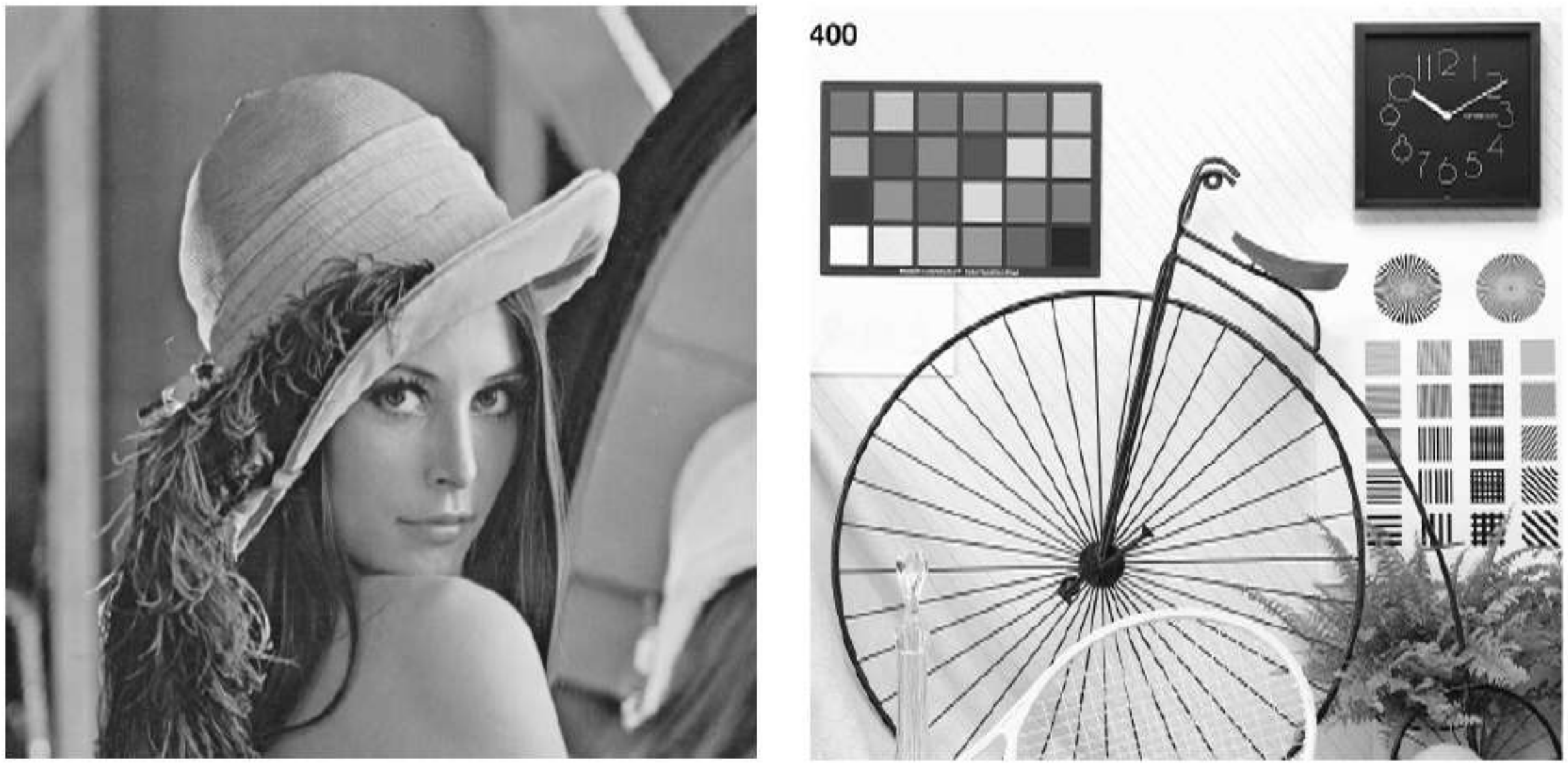

(a) (b)

**Fig 10** (a) Best Case Reconstruction. PSNR = 39.71. (b) Worst Case Reconstruction. PSNR = 24.45.

Normalized Size, PSNR for reconstruction and On-chip compression.

$$On-chip\ Compression = \frac{(16 - Bitdepth)}{16}, \tag{27}$$

where, $Bitdepth$ is the number of bits required to represent each pixel of compressively sampled image. Since each raw-pixel is 8 bits, and we are adding 2 pixels while doing CS, we are calculating on-chip compression relative to 16 bits in Eq. (27).

Similarly, Table 3 shows the same for the non-binary matrix. The best case and worst case image reconstruction for the non-binary CS followed by lossless JPEG is shown in Fig. 10.

**Table 1** Result for Baseline Image.

| $ImageType$ | $Quality$ | $Bitdepth$ | $Normalized\ Size$ | $PSNR(dB)$ |
|---|---|---|---|---|
| $JPEG$ | 75 | 8 | 100 | 34.87 |

Since we are adding two $8$ bit pixels for the binary sampling matrix, we need $9$ bits to represent the addition perfectly. For the non-binary matrix, we are using weights of $9$ and $7$ for each pixel. So the max value of weighted pixels can be $16 \times 255$. Hence we need $12$ bits to represent the weighted addition perfectly.

**Table 2** Results for Binary Block Diagonal matrix.

| $ImageType$ | $Quality$ | $Bitdepth$ | $Normalized\ Size$ | $PSNR(dB)$ | $On-Chip\ Compression(\%)$ |
|---|---|---|---|---|---|
| $JPEG + CS_B$ | $lossless$ | 9 | 220.42 | 32.45 | 43.75 |
| $JPEG + CS_B$ | $lossless$ | 8 | 188.05 | 32.43 | 50 |
| $JPEG + CS_B$ | $lossless$ | 7 | 154.99 | 32.37 | 56.25 |
| $JPEG + CS_B$ | $lossless$ | 6 | 121.97 | 32.19 | 62.5 |
| $JPEG + CS_B$ | $lossless$ | 5 | 92.58 | 31.58 | 68.75 |
| $JPEG + CS_B$ | 100 | 9 | 235.63 | 32.45 | 43.75 |
| $JPEG + CS_B$ | 100 | 8 | 198.19 | 32.44 | 50 |
| $JPEG + CS_B$ | 100 | 7 | 161.13 | 32.30 | 56.25 |
| $JPEG + CS_B$ | 100 | 6 | 126.23 | 31.93 | 62.5 |
| $JPEG + CS_B$ | 85 | 9 | 99.59 | 31.70 | 43.75 |
| $JPEG + CS_B$ | 85 | 8 | 69.55 | 30.87 | 50 |
| $JPEG + CS_B$ | 75 | 9 | 76.37 | 31.12 | 43.75 |
| $JPEG + CS_B$ | 75 | 8 | 51.68 | 30.02 | 50 |
| $JPEG+ CS_B$ | 75 | 7 | 33.75 | 28.64 | 56.25 |
| $JPEG + CS_B$ | 75 | 6 | 21.08 | 27.11 | 62.5 |
| $JPEG + CS_B$ | 75 | 5 | 12.33 | 25.45 | 68.75 |
| $JPEG + CS_B$ | 60 | 9 | 59.11 | 30.42 | 43.75 |
| $JPEG + CS_B$ | 50 | 9 | 51.97 | 30.03 | 43.75 |
| $JPEG + CS_B$ | 40 | 9 | 45.31 | 29.61 | 43.75 |
| $JPEG + CS_B$ | 30 | 9 | 38.08 | 29.05 | 43.75 |
| $JPEG + CS_B$ | 20 | 9 | 29.23 | 28.21 | 43.75 |

We can see from the Table 2 and Table 3 that the performance of the binary and non-binary matrix for CS with lossless JPEG and without bit-truncation is almost the same. This is in agreement with the results stated in Ref. [5]. We can also see that the storage size is very high for CS with lossless JPEG. This has the potential to degrade the performance of imaging system when comes to storage and we will need a much more complicated JPEG decoder. To decrease the size, we can either decrease $quality$ or truncate LSB's or both. By truncating LSB's we not only decrease the size of the image but also significantly simplify ADC design as well as JPEG encoder and decoder design. This simplified decoder will also consume less energy because of reduced switching activity resulting from reduced bitwidth. Similarly we can also decrease the quality factor to decrease the size. For example, if we use default Quality factor i.e. 75 we can see that performance loss is not much but size is much smaller.

In general, for a given quality factor, the non-binary matrix performs quite better than the binary matrix. This is because it can preserve much more information than binary matrix because of larger bitwidth. This makes it more resilient to degradation during JPEG quantization step. This is also evident in the graph shown in Fig. 11 where none of the LSB's have been truncated. The better PSNR for non-binary sampling matrix comes at the cost of increased image size. A comparison between the normalized image-size resulting binary and non-binary sampling matrix for $bitdepth = 9$ and $bitdepth = 12$ respectively is shown in Fig. 12. By pruning some LSB's we can decrease image size at the cost of the PSNR of reconstructed image. Thus the non-binary sampling matrix offers more control over image quality than the binary sampling matrix.

We can also see from Table 2 and Table 3 that for a given quality factor as we truncate the LSB's of CS sampled image in the non-binary sampling method, the result approaches that of the binary sampling method i.e. the performance of the non-binary matrix almost equals that of the binary matrix for same bitdepth. For the maximum performance case i.e. CS with lossless JPEG, the performance of both sampling matrix is same for full bitdepth for each case respectively. While the result for maximum performance case for CS is roughly 2dB less than the basline JPEG case of Table 1, the former provides roughly $43\%$ raw data compression but the latter provides none. Reduction in raw data rate will significantly simplify our system design. This is discussed in our next section.

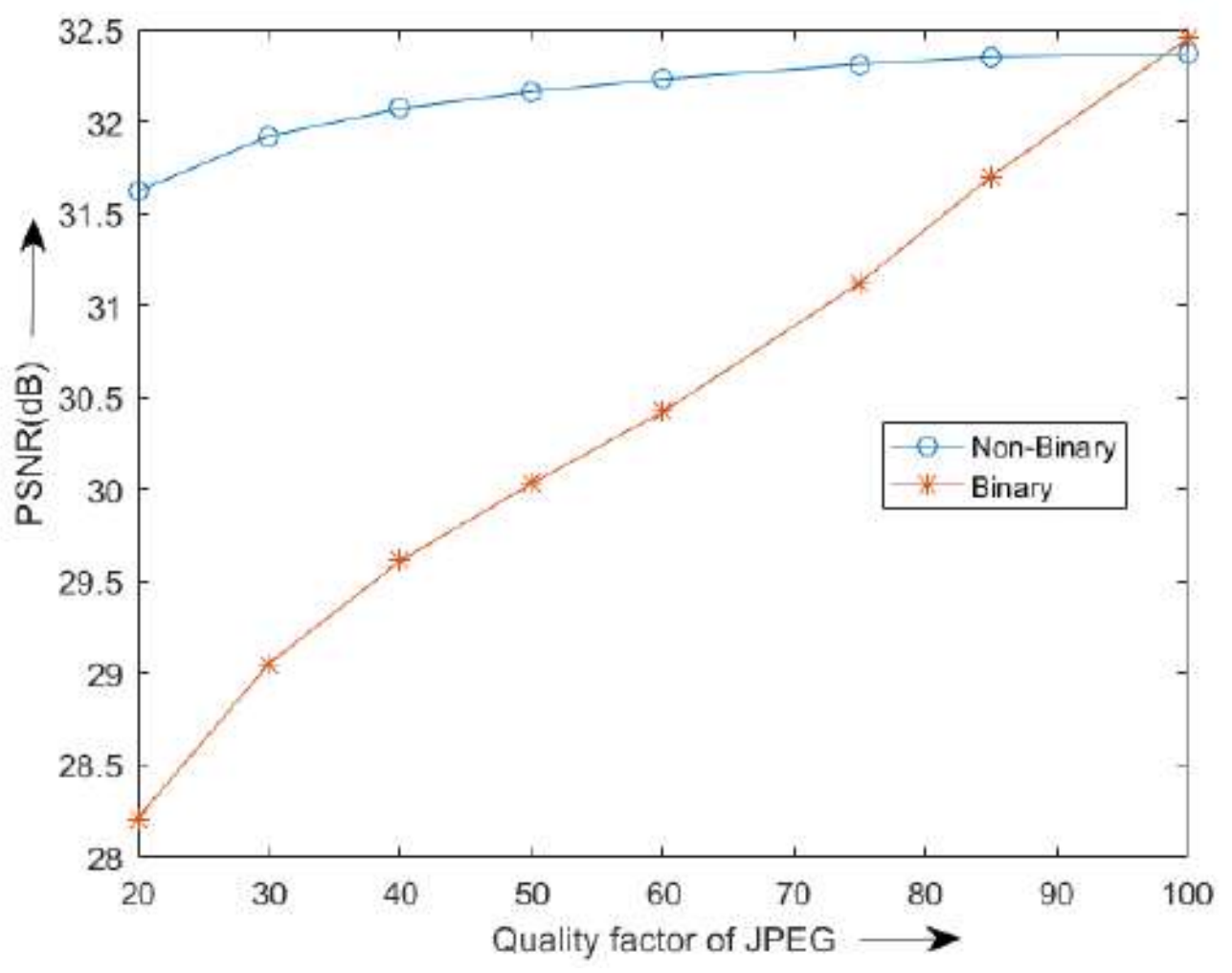

**Fig 11** Graph showing PSNR of image reconstruction for binary and non-binary matrix cases vs JPEG Quality. LSB's have not been truncated.

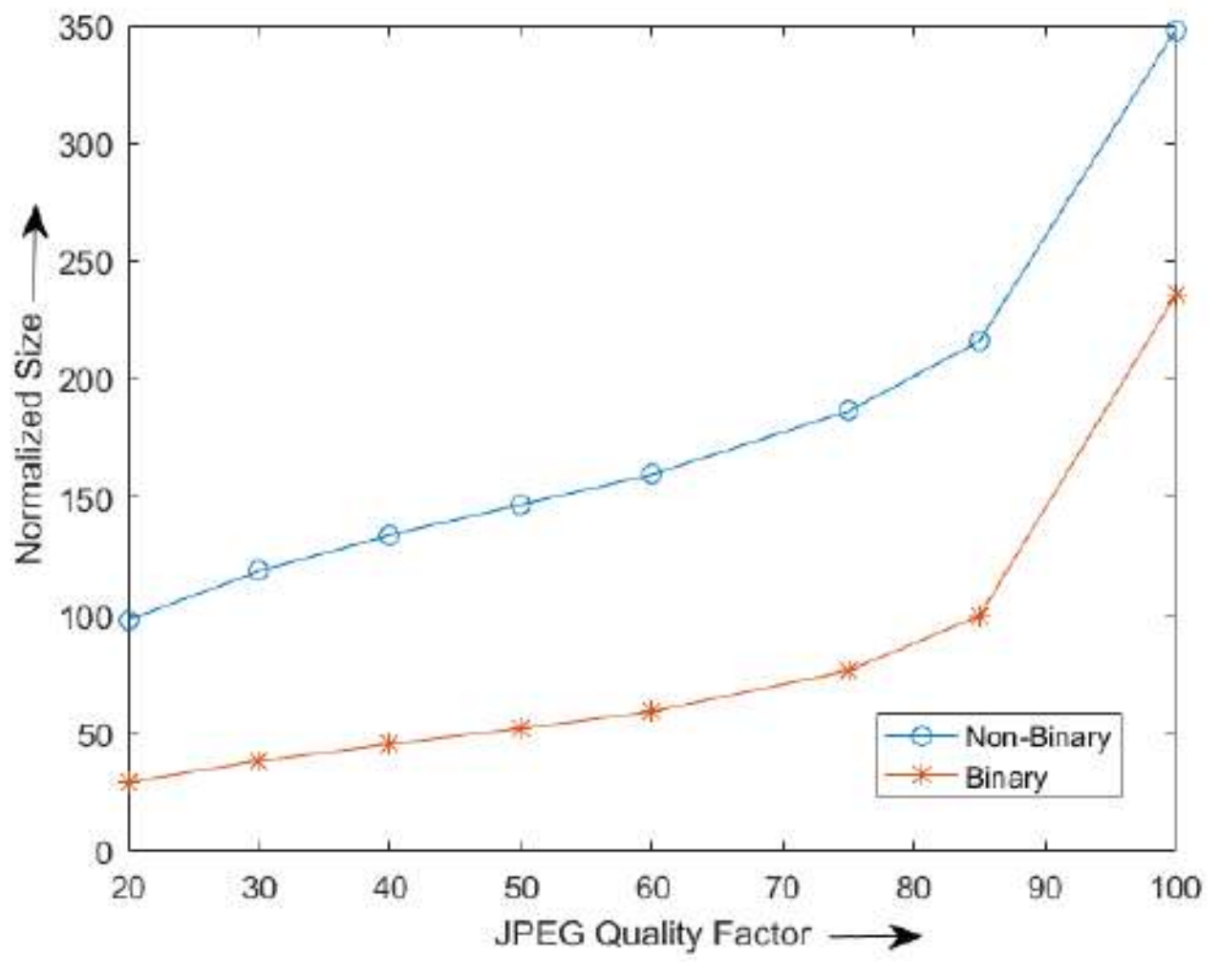

**Fig 12** Graph showing normalized image-size binary and non-binary matrix cases vs JPEG Quality. LSB's have not been truncated.

These were the simulations for gray-scale images. For colored images the procedure is straightforward. In the case of RGB image, the three different color planes can be thought of as three different images and CS can be applied to each of the 3 images. The reconstruction performance for colored Lenna image is mentioned in Table 4.

The next section will discuss the novel implementation of front-end sampling matrix on image sensor level.

## 4 Design

This section discusses the novel sensor level design to implement front-end sampling matrix presented in the previous section. It also discusses briefly about the ADC and JPEG encoder.

When comes to hardware implementation, binary block diagonal matrix means an addition of the row or column pixels. The number of pixels to be added is the number of ones in the row of sampling matrix. For our binary sampling matrix, we can simply implement this by using double sized pixels. We can choose any pixel design i.e. 3T or 4T. Large pixels have better SNR values because dark current decreases much faster than sensitivity as area increases (Ref. [24]). Even if noise is larger in smaller pixels, it is taken care of by using Correlated Double Sampling technique. So the higher noise level of smaller pixel is not much of an issue. If we use a large photodiode to implement binary sampling matrix, it means increase in the fill-factor of pixel. If fill factor for a given pixel design is $f$ then using a double sized photodiode will roughly give $2f/(1+f)$ fill factor. For $f = 0.7$ we get a rough approximation for new factor as $f = 0.82$. This increased fill factor can compensate the loss due to reconstruction algorithm.

The non-binary block diagonal matrix has to perform weighted addition. This can be done by using our novel design shown in Fig. 13. This is inspired by the 4T design. We have used a very simple technique to perform weighted addition. We have used a small capacitance (gate capacitance of MOS) to decrease response of one of the photodiode by placing it before the shutter or Tx Transistor. This MOS is labeled as $cap$ in Fig. 13. This effectively decreases the sensitivity of the photodiode and it generates less output ( output of a photodiode is actually a decrease in the output voltage w.r.t. reset voltage level of photodiode because photocurrent flows to discharge the junction capacitance of photodiode) as compared to the other photodiode without additional capacitance. So if the same amount of light falls in both photodiode then one photodiode will generate less output voltage than the other. When the shutter MOS (i.e. Tx_1 and Tx_2) opens, then current drains from the floating diffusion node to the photodiode. Since one photodiode has less voltage than other so one will draw less current than other. This is because our circuit is operated in transient state rather than steady state. The shutter open time is set such that the circuit remains in transient state. Since both the currents are unequal, the resulting voltage at the floating diffusion node i.e. FD is like weighted addition of two equal signals. For the non-binary sampling matrix, we used weights of $9$ and $7$. So, the relative weight of one pixel w.r.t. to another is approximately $1.3(9/7)$. The circuit depicted in Fig. 13 also achieves approximately the same weight. Since even after truncating some LSB's we can get good images, the weighted addition does not have to be very exact as the errors will get truncated too. The Spectre simulation results for the circuit are stated in Table 5. The weight has been calculated in the table keeping CDS technique in mind. The weight has been calculated by curve fitting for $100$ different points. For generating these points, the photocurrent in each photodiode was varied from $100fA$ to $1000fA$ in steps of $100fA$. This generates $10$ points for each photodiode. Then all possible permutations of these $2$ sets (one set for each photodiode) of $10$ points were taken to generate $100$ different points. Fig. 14 shows the sweep analysis performed for these 100 points (offset voltage has been removed). Fig. 15 shows a plot to demonstrate weighted addition of photodiode outputs. The curves in the plot represents the output voltage values for the proposed pixel circuit for $2$ different cases. In each case, the photocurrent of one of the photodiode is fixed at $100fA$ and other one is varied form 100fA to 100fA in steps of 100fA. Thus, for a given current value in x-axis of the plot, total charge generated in the pixel will be same. But, the output of pixel is different for both cases because of weighted addition of photodiode output.

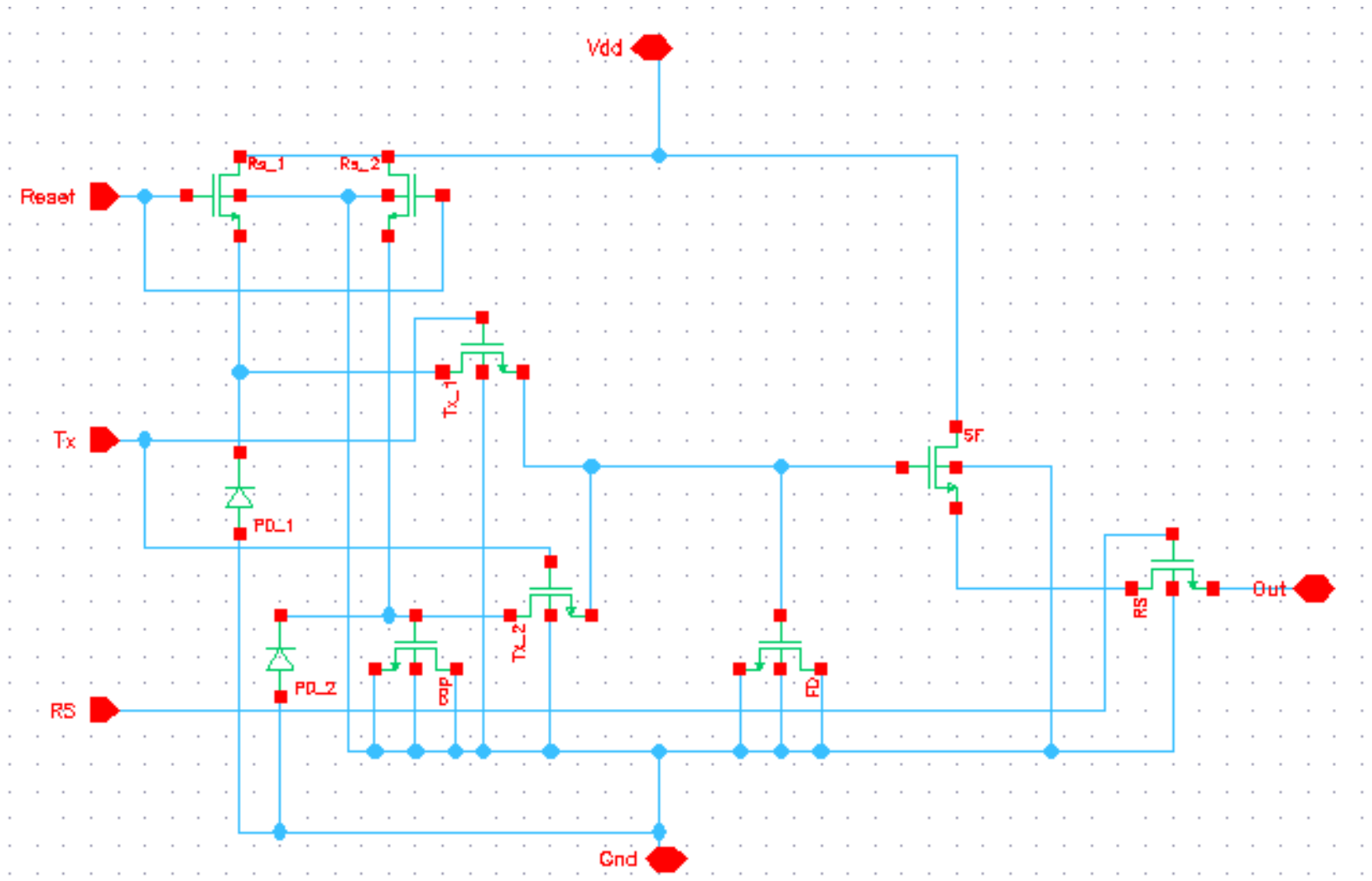

**Fig 13** Schematic design for on-pixel compressed sensing.

Addition of capacitor in one of the photodiode results in a decrease on sensitivity. In traditional designs, a decrease of sensitivity implies a loss of resolution, but in our design reconstruction algorithms help us recover this information.

If we are truncating the bits, we are significantly simplifying ADC design too. Bit truncation in the simulation can be implemented in hardware by decreasing the ADC resolution. This will result in a simpler and power efficient ADC. Since at lower resolutions noise and linearity requirements are relaxed, voltage scaling can help us achieve an exponential reduction in power consumption (Ref. [27]). Since ADC is responsible for a major chunk of power consumption during the process of raw image acquisition (Ref. [1, 28]), our technique will have a significant impact in reducing the power consumption.

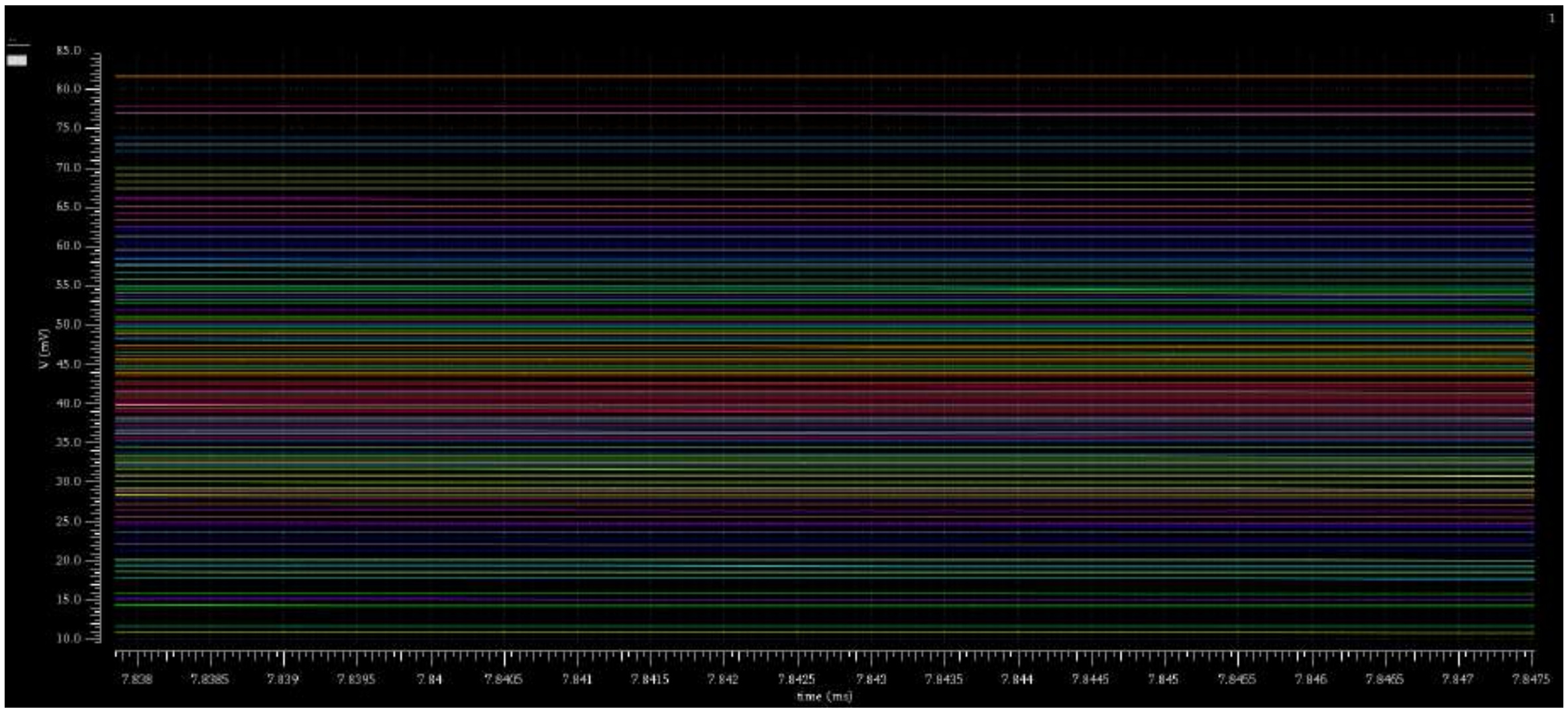

**Fig 14** Sweep analysis for our proposed pixel circuit.

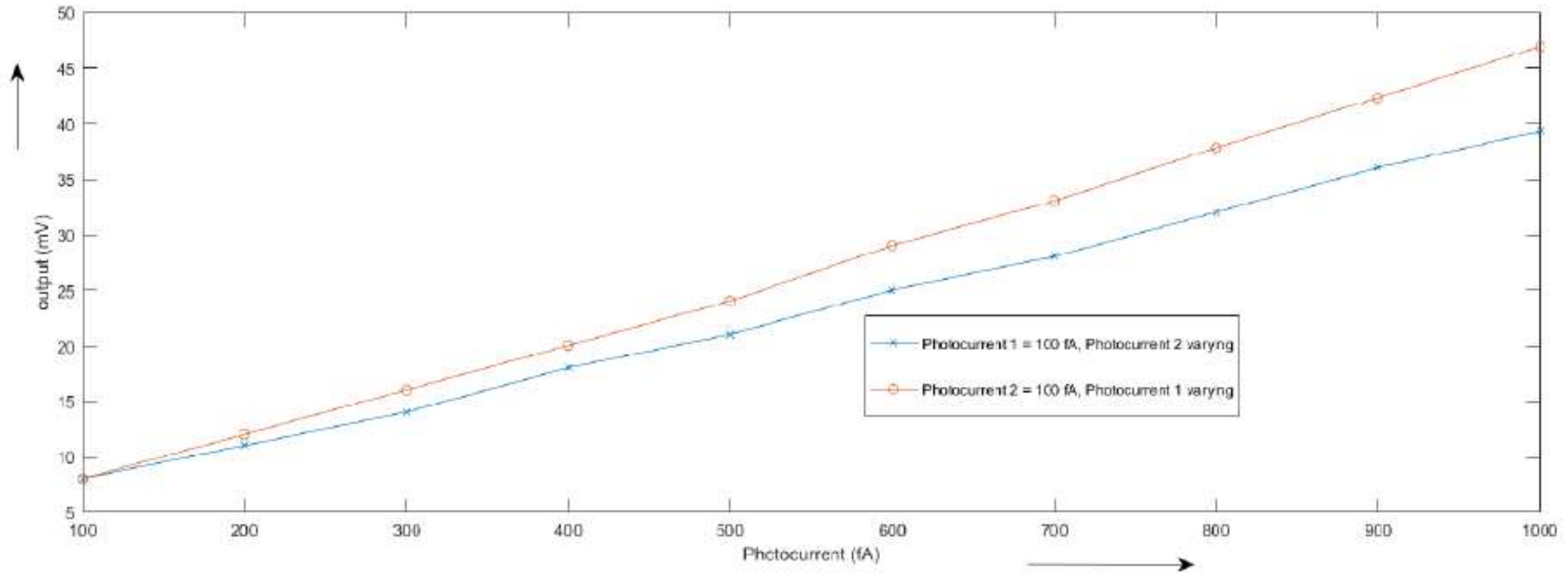

**Fig 15** Plot showing weighted addition of photodiode outputs. For each curve, photocurrent in one of the photodiode is fixed at 100fA while the other one is varied from 100fA to 1000fA in steps of 100fA. Each point in x-axis represents same amount of charge generated in the pixel but output is different due to weighted addition of photodiode outputs.

We have designed our pixel for both Front Side Illumination and Backside Illumination (BSI) (Ref. [29]). The FSI layout for the Fig. 13 circuit is shown in Fig. 16. In FSI layout light enters from the frontside of the sensor where as in BSI it enters from the backside. This means in BSI we can draw metal lines over photodiode and increase fill factor. There are two different technologies in BSI which are shown in Fig. 17. They are conventional BSI and stacked BSI (Ref. [29]). In conventional BSI, the logic circuit and the pixel circuit are in the same plane. Metal wiring can be drawn over pixel circuit as light enters from backside. This results in an increase of the fill factor. In stacked BSI, the logic circuit and pixels are in different planes. This means the fill factor is almost 100% for stacked BSI. The layout for conventional BSI and stacked BSI for our novel pixel circuit is given in Fig. 18 and Fig. 19. We used TSMC 200nm technology library and Cadence Design tools to implement our design. The advantages associated with on-chip implementation of CS does not depend on the technology of choice. It works equally well in any technology.

The junction capacitance, responsivity and dark current for the photodiode used in our pixel was estimated using the data and graphs presented in Ref. [24] and Ref. [23]. The formula for junction capacitance is given below,

$$C_{jdep} = \frac{C_{J0}A_D}{1-(\frac{V_d}{v_j})^m} + \frac{C_{J0sw}P_D}{1-(\frac{V_d}{v_{jsw}})^{m_{jsw}}} \tag{28}$$

where $C_{J0}$ and $C_{J0sw}$ represent zero-bias capacitance at the bottom and sidewall components, $V_d$ is the voltage applied to the photodiode, $v_j$ and $v_{jsw}$ stand for the built-in potential of the bottom and the sidewall respectively, $m$ and $m_{jsw}$ are the grading coefficients of the bottom and the sidewalls, $A_D$ is the photodiode area in $m^2$ and $P_D$ represents the photodiode perimeter in $m$. These parameters are given in the Table 6 for our design. The table also lists the fill factor for our layout.

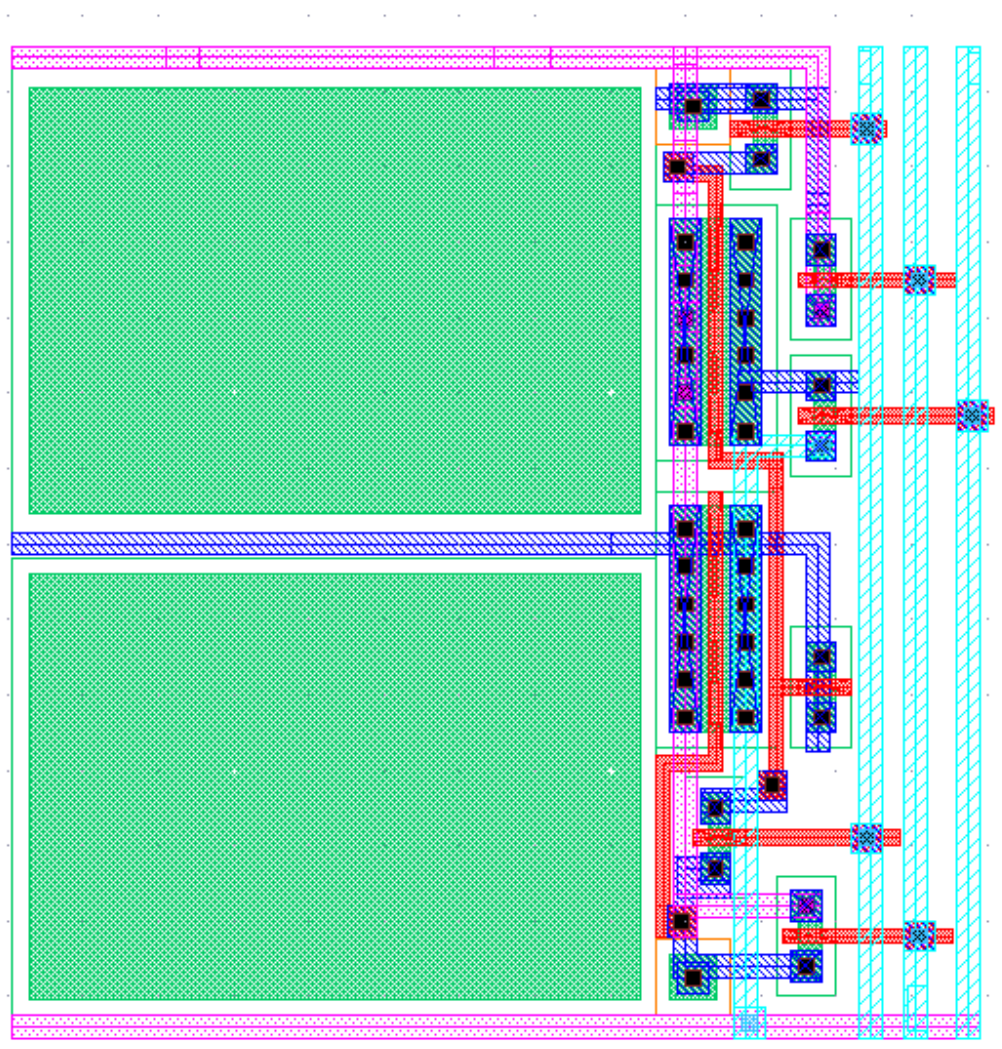

**Fig 16** Proposed FSI Pixel Layout to implement on-chip compressed sampling.

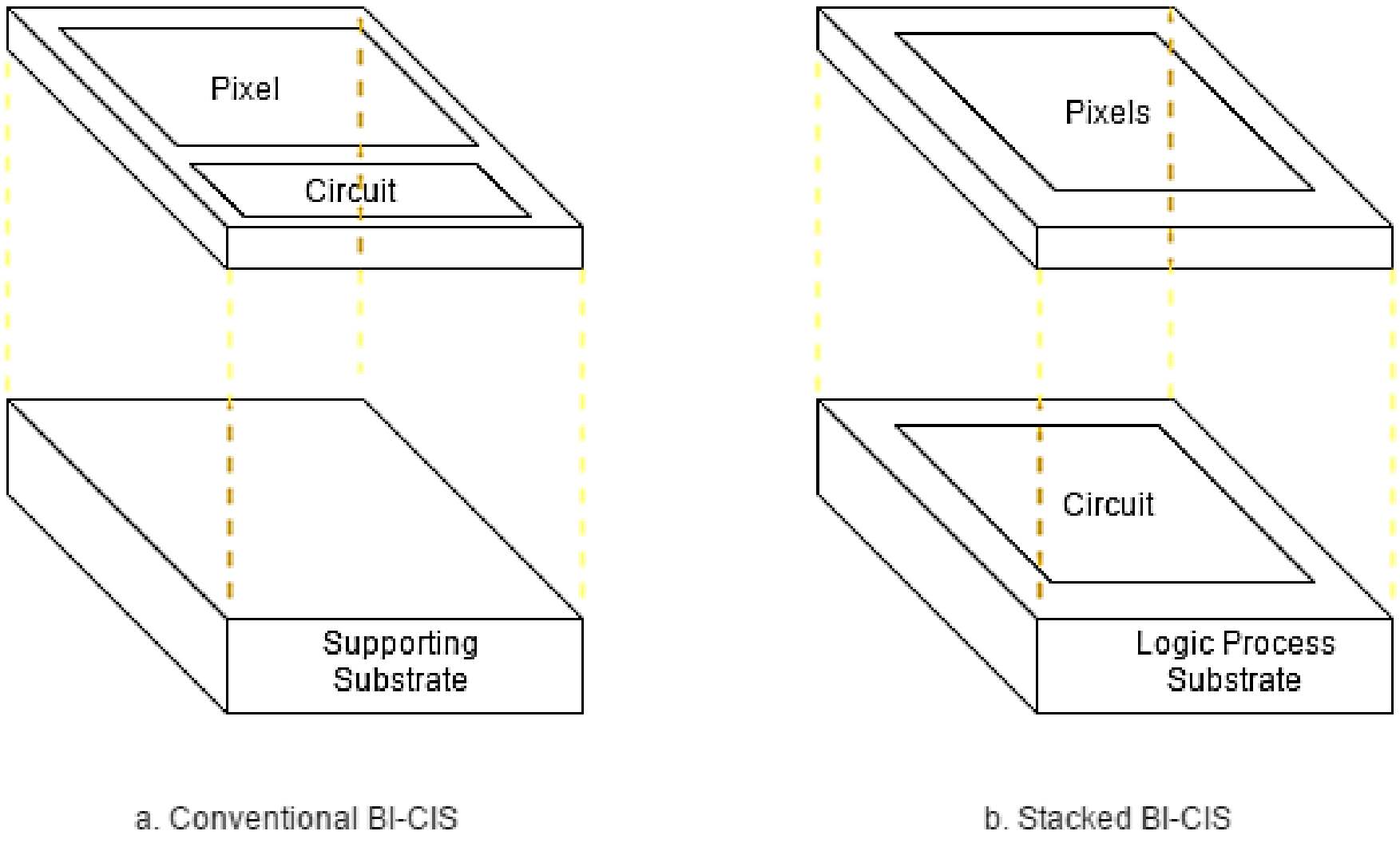

**Fig 17** A schematic diagram explaining different Back-illuminated CMOS Image Sensor (BI-CIS).

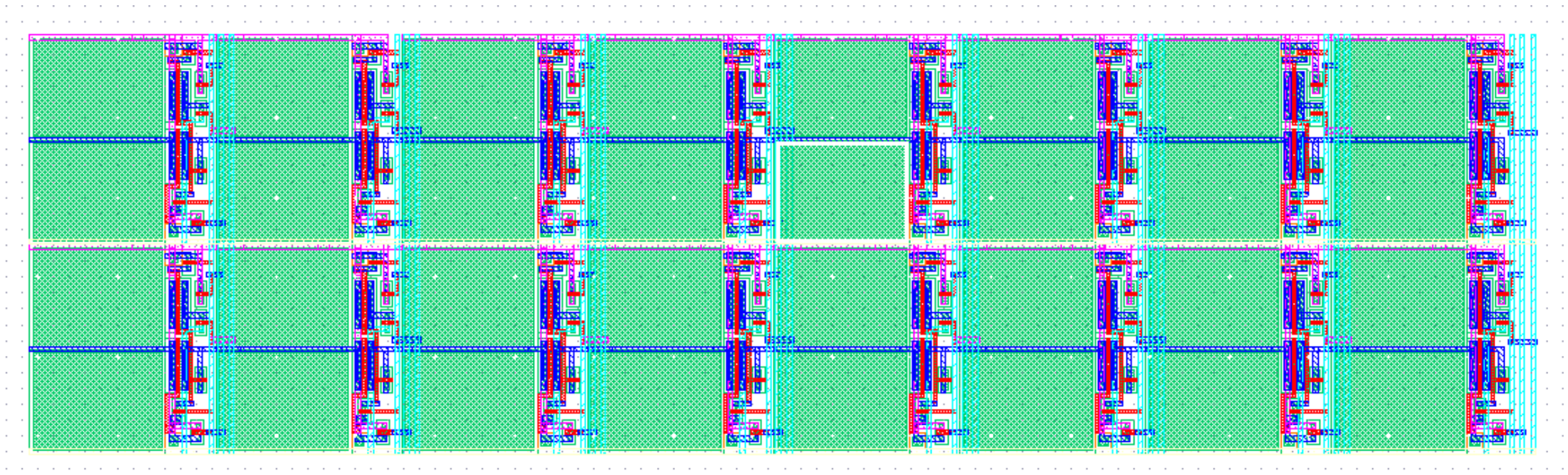

**Fig 18** Proposed Conventional BSI Layout for pixels.

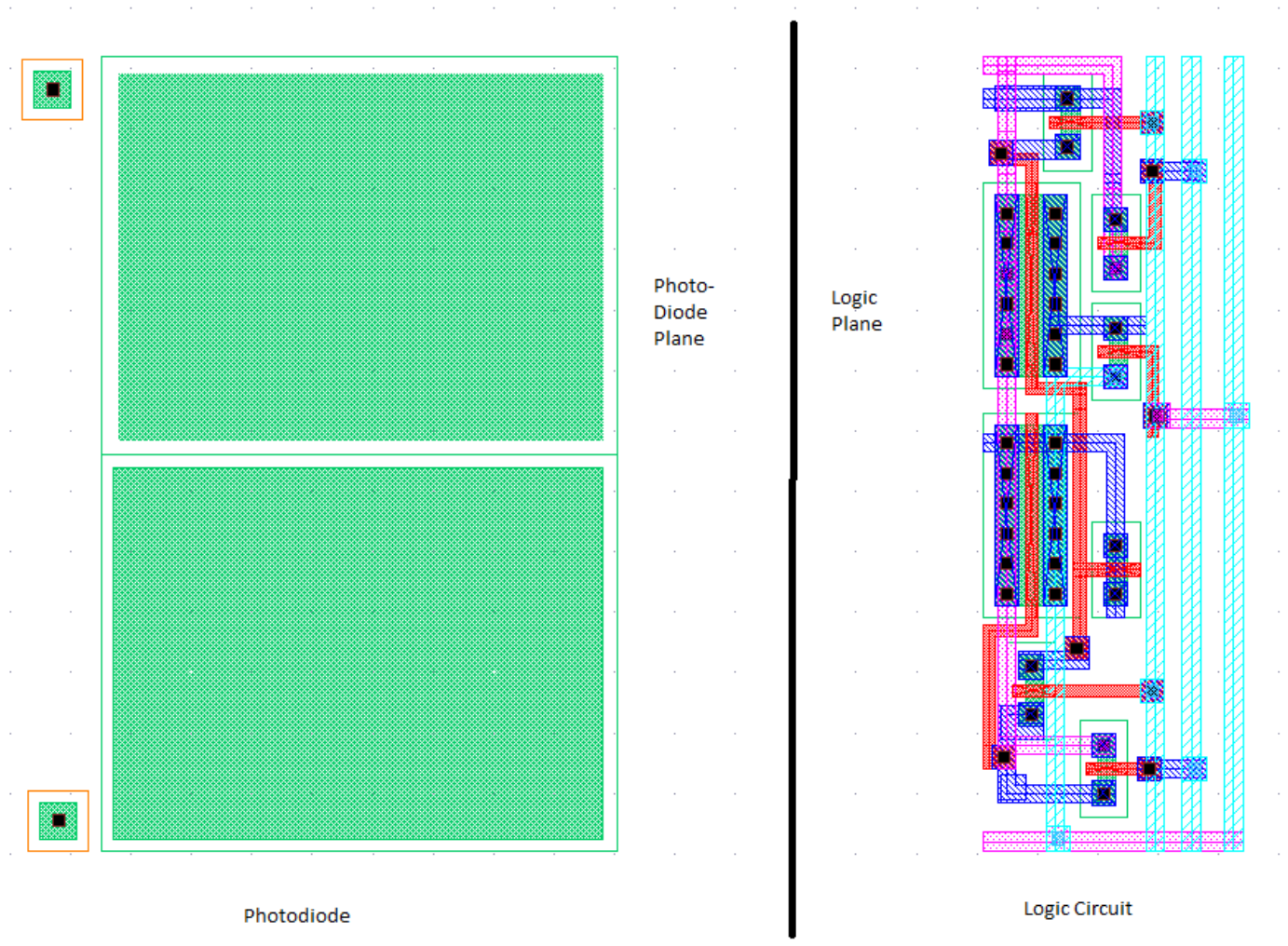

**Fig 19** Proposed Stacked BSI Layout for pixels.

A problem with our pixel design is that it is non-linear. Switching transistor, source follower, active capacitances, all contribute to non-linearity. This non-linearity can be removed by curve fitting. In an actual system a lookup table can be used to simplify implementation. The equation obtained after curve fitting is reproduced below -

$$output = 1.037 - (3.324 \times 10^{-5})p2 - (4.065 \times 10^{-5})p1 - (1.678 \times 10^{-9})p2^2 \\ -(3.88 \times 10^{-9})p1p2 - (2.564 \times 10^{-9})p1^2, \tag{29}$$

where $p1$ and $p2$ refers to photocurrent in photodiode 1 and photodiode 2 respectively in $fA$ and output refers to output voltage in $V$ of the pixel shown in Fig. 13. The weight for wighted addition has been calculated by taking the ratio of coefficients of $p1$ and $p2$.

Our system design methodology simplifies JPEG encoder and decoder design as well. JPEG generally takes DCT of image blocks of size $8 \times 8$. Since we are combining two pixels to one we are effectively reducing the number of blocks by half. This will cut energy spent during encoding by half. Since the encoder design is mostly pipelined, it will also reduce encoding latency by half. Since workload is reduced one can reduce the voltage and frequency of operation of JPEG encoder to maintain same latency. This will result in an exponential decrease in energy consumption during encoding and decoding process. If we truncate the LSB's of image this will lead to additional simplification of encoder and decoder design and power savings. It leads to a proportional decrease in switching activity and hence dynamic power. It reduces register as well as arithmetic unit bitwidth. Reduction in bitwidth of arithmetic unit can lead to a direct reduction in latency of such units and the chip floor-area.

Other implementations for both the binary and non-binary sampling matrix is also possible. For implementing the binary sampling matrix we can also use the design presented in Fig. 13. We can remove the $cap$ MOS from the circuit to do so. This design will be especially useful when we have two photodiodes separated apart, as in colored image sensor implemented using popular Bayer Pattern (Ref. [30]). This is shown in Fig. 20. One can see that each photodiode representing a color have to be separated apart. It might not be possible to make large single color photodiode because of resolution reasons. Hence we can use design presented in Fig. 13 for both the non-binary matrix as well as binary matrix (i.e. without $cap$).

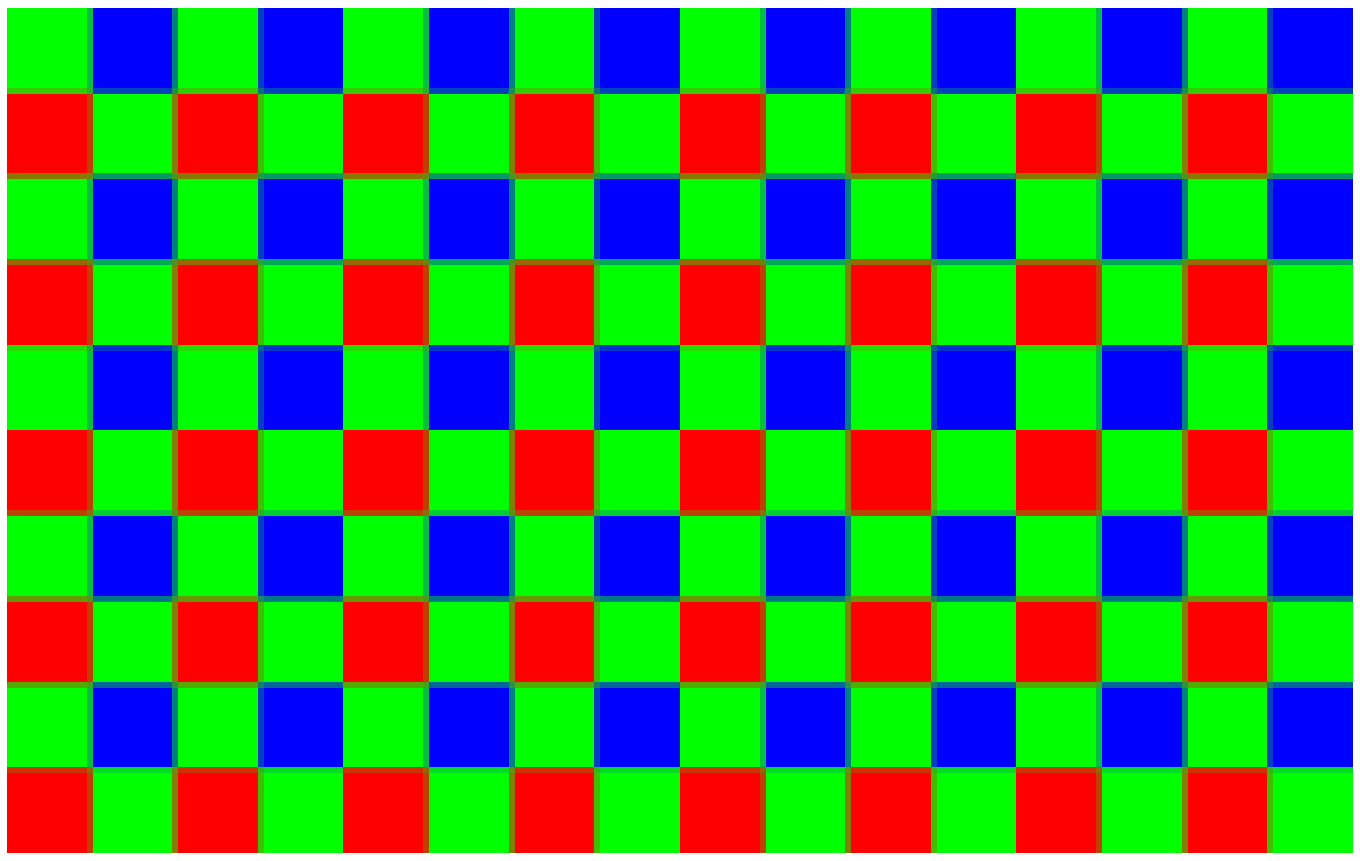

**Fig 20** Bayer Pattern

While the theoretical assumption is that all source followers or photodiode provide same gain/response, this is hardly the case practically. Because of manufacturing inconsistencies, the two photodiodes will have different responses and parasitics. So the binary matrix will become non-binary in actual implementation. This will not pose any problem because non-binary matrix works equally well. By incorporating such inconsistencies further into the sampling matrix, we can solve problems posed by fixed pattern noise during the reconstruction. There are multiple sources for fixed pattern noise. Photoresponse non-uniformity, source follower mismatch etc. are sources of mismatch (Ref. [31]). We can handle this mismatch by incorporating the mismatch in the sampling matrix. If all source followers or photodiodes provide different gain/response then we can use different weights for our sampling matrix for each of the pixels. If we truncate some LSB's, then we do not need incorporate mismatch also as long as noise is less than or equal to the LSB's truncated. This is because noise will be truncated with the LSB's. For CMOS image sensor design presented in Ref. [28], the FPN noise is less than 1 LSB. So if we implement CS with bit truncation in such system, we can get rid of the noise by truncating just 1 LSB.

Yet another way to implement binary or non-binary matrix is to sum the pixels at Analog to Digital Converter (ADC) level, similar to what was presented in Ref. [1]. This has an advantage of having an option to choose between CS mode and non-CS mode of operation. But one has to pass address for each pixel. This will slow down frame rate and increase power consumption. There are certain additional disadvantages associated with it which are discussed later in the paragraph below.

The advantage of CS implemented at the sensor level is not just limited to reduction of data rate only. Our implementation shown in Fig. 13 for the non-binary matrix requires only 6 transistors (excluding floating diffusion nodes and capacitance) per 2 pixels i.e. 3 transistors per pixel with global shutter. This means improvement in fill factor and reduction in the size of pixels. This also means less power consumption. A simple analysis of power consumption for image acquisition can be performed by looking at the data mentioned in Ref. [1] and Ref. [28]. Both the papers use different design, technology and specification. Hence power consumption is different for both of them. But the relative breakdown of power spent in I/O, ADC, Pixel and Other operations are approximately the same. Roughly 90% power is spent in I/O and ADC in the image acquisition. Our proposed CS implementation cuts the I/O and ADC operations exactly by on-chip compression ratio i.e. by $25\% - 68.75\%$. We have used the data for the normal mode of operation at 120 frames/sec in our work. The data has been reproduced in Table 7. The table also lists the estimation of power if our system design methodology is implemented in the normal image sensor described in the paper. We have also incorporated the power spent during JPEG compression in the same table using the design presented in Ref. [32] as reference. Please note that we have only performed rough approximation of JPEG power consumption based switching activity. We can see from Table 7 that one can achieve approximately $23.5\% - 65\%$ power savings for compression ratio of $25\% - 68.75\%$ respectively. In this work, we only consider the energy spent during image acquisition and compression process.

Our proposed CS technique also reduces the wiring area in the die as we combine two rows/columns in one. We cut the amount of wiring required by half. We need half the number of rows or column select, Reset and Transmit for Global Shutter. We also need half size address decoder leading to a reduction in power and chip area. Because of the reduced size of the pixel and reduced wiring area, we can fit more pixels in the same die area using existing technology. It is not possible to exploit these advantages if we implement CS at ADC level as mentioned previously and in Ref. [1].

## 5 Conclusions

We have discussed a novel system design methodology for an imaging system which significantly cuts down power and simplifies hardware design. By using simple deterministic matrices presented in this paper, CS can be used for on-chip compression of the raw image to save power. These matrices also helps us to use JPEG in conjunction with CS for additional compression. We have also presented novel pixel design implementing such matrices and results for on-chip compression ranging from $25\% - 68.75\%$. This leads to significant reduction in power spent at I/O, ADC and JPEG. We can significantly simplify ADC design as we require less resolution and less speed. Similarly we can simplify JPEG encoder design because there are half the number of $8 \times 8$ image blocks and reduced bitwidth per pixel. When we do not use voltage-frequency scaling, we save the power by approximately the same amount as on-chip compression. In such case estimated power savings are around $23.5\% - 65\%$ for the on-chip compression ratio of $25\% - 68.75\%$ respectively. If we use scaling we can achieve exponential reduction also. We also require fewer transistors to implement the same number of pixels. For the design proposed in Fig. 13 we need three transistors per pixel to implement global shutter which in normal case demands four. This not only increases the fill factor of pixels but decreases the pixel size as well. We also reduce the amount of wiring required by half which means a significant reduction in crosstalk and pixel size. We need only half the number of row/column wiring, power supply, reset and global shutter control wiring. We also simplify row/column address decoder significantly as we need only half the size decoder. All this means we can fit more pixels in a given die area while maintaining power efficiency and more than compensate the loss due to reconstruction algorithm. Since amount of raw data is reduced, we can also increase the frame rate. Thus our system design methodology has a huge potential to increase the power efficiency of CMOS image sensor design while increasing resolution and significantly simplifying circuit design. This aspect should be explored further by testing prototypes.

## 6 Future Work

This paper presents a novel methodology for an image acquisition system. The performance of this methodology depends on the performance of each component of this system. Each component can be redesigned to suit the needs of this system. In particular, the image sensor presented in this paper can be fabricated and tested to get more accurte results. Another significant component of this system is the reconstruction algorithm. An improvement in the performance of reconstruction algorithm can significantly improve the performance of this system. Post fabrication, the noise characteristics of the system should also be studied and improved upon.

*Acknowledgments*

The authors would like to thank Sushirdeep Narayana for his valuable inputs.

**Pravir Singh Gupta** received his B.Tech in Electrical Engineering from Indian Institute of Technology Bhubaneswar, India in 2013. He is doing his PhD in Electrical and Computer Engineering Department at Texas A & M University at College Station. He has interests in on-sensor image compression, LDPC decoders and Polar Decoders.

**Gwan Seong Choi** received his B.S., M.S. and Ph.D. degrees, all in electrical and computer engineering, from the University of Illinois at Urbana-Champaign in 1988, 1989 and 1994, respectively. He currently is an associate professor in the Electrical and Computer Engineering department at Texas A& M University. He has worked for Cray Research Inc. and Tandem Computers Inc, and he has been a visiting scientist at the NASA Langley Research Center. Dr. Choi's research interests include fault-tolerance, verification simulation, high-performance VLSI circuits, radiation testing, design for dependability and software engineering.

# List of Figures

# List of Tables

## 7 Table for Power Estimation

**Table 3** Results for Non-Binary Block Diagonal matrix.

| $Image\,Type$ | $Quality$ | $Bitdepth$ | $Normalized\,Size$ | $PSNR(dB)$ | $On-Chip\,Compression(\%)$ |
|---|---|---|---|---|---|
| $JPEG + CS_{NB}$ | $lossless$ | 12 | 317 | 32.37 | 25 |
| $JPEG + CS_{NB}$ | $lossless$ | 11 | 285.57 | 32.37 | 31.25 |
| $JPEG + CS_{NB}$ | $lossless$ | 10 | 253.55 | 32.37 | 37.5 |
| $JPEG + CS_{NB}$ | $lossless$ | 9 | 221.88 | 32.37 | 43.75 |
| $JPEG + CS_{NB}$ | $lossless$ | 8 | 189.23 | 32.36 | 50 |
| $JPEG + CS_{NB}$ | $lossless$ | 6 | 122.27 | 32.13 | 62.5 |
| $JPEG + CS_{NB}$ | $lossless$ | 5 | 92.82 | 31.52 | 68.75 |
| $JPEG + CS_{NB}$ | 100 | 12 | 348.21 | 32.37 | 25 |
| $JPEG + CS_{NB}$ | 100 | 11 | 310.88 | 32.37 | 31.25 |
| $JPEG + CS_{NB}$ | 85 | 12 | 216.14 | 32.35 | 25 |
| $JPEG + CS_{NB}$ | 85 | 11 | 176.32 | 32.29 | 31.25 |
| $JPEG + CS_{NB}$ | 75 | 12 | 186.45 | 32.31 | 25 |
| $JPEG + CS_{NB}$ | 75 | 11 | 146.28 | 32.16 | 31.25 |
| $JPEG + CS_{NB}$ | 75 | 10 | 108.39 | 31.78 | 37.5 |
| $JPEG + CS_{NB}$ | 75 | 9 | 76.48 | 31.09 | 43.75 |
| $JPEG + CS_{NB}$ | 75 | 8 | 51.78 | 29.99 | 50 |
| $JPEG + CS_{NB}$ | 75 | 7 | 33.79 | 28.63 | 56.25 |
| $JPEG + CS_{NB}$ | 75 | 6 | 21.11 | 27.11 | 62.5 |
| $JPEG + CS_{NB}$ | 60 | 12 | 159.67 | 32.23 | 25 |
| $JPEG + CS_{NB}$ | 50 | 12 | 146.85 | 32.16 | 25 |
| $JPEG + CS_{NB}$ | 50 | 10 | 76.86 | 31.09 | 37.5 |
| $JPEG + CS_{NB}$ | 50 | 9 | 52.05 | 30.01 | 43.75 |
| $JPEG + CS_{NB}$ | 40 | 12 | 133.90 | 32.07 | 25 |
| $JPEG + CS_{NB}$ | 40 | 10 | 67.86 | 30.78 | 37.5 |
| $JPEG + CS_{NB}$ | 30 | 12 | 118.58 | 31.92 | 25 |
| $JPEG + CS_{NB}$ | 20 | 12 | 97.93 | 31.62 | 25 |

**Table 4** Results for Non-Binary and binary block diagonal matrix for colored Lenna image.

| $CS\ Type$ | $PSNR(dB) - Red$ | $PSNR(dB) - Green$ | $PSNR(dB) - Blue$ |
|---|---|---|---|
| $Lossless\ JPEG + CS_B$ | 41.27 | 37.52 | 35.94 |
| $Lossless\ JPEG + CS_{NB}$ | 41.22 | 37.48 | 35.91 |

**Table 5** Results of Spectre simulation for weight calculation.

| $Photodiode\ 1\ (fA)$ | $Photodiode\ 2\ (fA)$ | $Output(Pixel\ Circuit)\ (mV)$ |
|---|---|---|
| 100 | 100 | 8 |
| 200 | 100 | 11 |
| 100 | 200 | 12 |
| 500 | 500 | 38.9 |
| 1000 | 1000 | 81.7 |
| $Calculated\ weight\ using\ curve\ fitting(100\ points)\ =\ 1.22$ | | |

**Table 6** Table for Photodiode and Pixel Parameters

| $C_{jdep}$ $(fF)$ | $C_{j0}$ $(mF/m^2)$ | $C_{j0sw}$ $(F/m)$ | $v_j$ $(V)$ | $v_{jsw}$ $(V)$ | $m$ | $m_{jsw}$ | $V_d$ $(V)$ | $A_D$ $(\mu m^2)$ | $P_D$ $\mu m$ | $FSI$ $F.F.$ | $BSI$ $F.F.$ |
|---|---|---|---|---|---|---|---|---|---|---|---|
| 32.8 | 1.067 | $1.6 \times 10^{-10}$ | .8 | .65 | .41 | .35 | 1.8 | 45.76 | 27.5 | 54.16% | 65.54% |

Note. F.F. stands for Fill-Factor of pixel. In case of BSI, Fill Factor is mentioned for conventional BSI only.

**Table 7** Table for Power Estimation

| **Operation** | **Design 1 (mW) (Ref. [1])** | **Design 2 (mW) (Ref. [28])** | **CS for Design 1 (mW)** $C.R.: 68.75\% - 25\%$ | **CS for Design 2 (mW)** $C.R.: 68.75\% - 25\%$ |
|---|---|---|---|---|
| I/O | $27$ | $70$ | $8.34 - 20.25$ | $21.87 - 52.5$ |
| ADC | $60$ | $209$ | $18.75 - 45$ | $65.31 - 156.75$ |
| Pixel | $1.8$ | $23$ | $1.8$ | $23$ |
| Other | $4.2$ | $20$ | $4.2$ | $20$ |
| JPEG (Ref. [32]) | $13.18$ | $386.3$ | $4.11 - 9.88$ | $120.71 - 289.72$ |
| Total | $106.18$ | $708.3$ | $37.2 - 81.13$ | $250.89 - 541.97$ |
| Power Savings | $0\%$ | $0\%$ | $64.96\% - 23.59\%$ | $64.57\% - 23.48\%$ |

Note. CS stands for our proposed Compressed Sensing and C.R. stands for on-chip Compression Ratio.